\pdfoutput=1
\documentclass[fleqn,usenatbib]{mnras}

\usepackage{newtxtext,newtxmath}

\usepackage[T1]{fontenc}
\usepackage{ae,aecompl}

\usepackage{graphicx}	
\usepackage{amsmath}	
\usepackage{amssymb}	
\usepackage{stmaryrd}
\usepackage{array,multirow}
\usepackage{bm}
\usepackage{subfig}
\usepackage{inputenc}
\usepackage[dvipsnames]{xcolor}

\title[A Pixel Space Method for Dipole Modulation]{A Pixel Space Method for Testing Dipole Modulation in the CMB Polarization}

\author[S. Ghosh and P. Jain]{
Shamik Ghosh,$^{1}$\thanks{E-mail: shamik@ustc.edu.cn} \thanks{Current Affiliation: Dept. of Astronomy, University of Science and Technology of China, Hefei - 230022, China}
and Pankaj Jain,$^{1}$\thanks{E-mail: pkjain@iitk.ac.in}
\\
$^{1}$Dept. of Physics, Indian Institute of Technology, Kanpur - 208016, India 
}

\date{Accepted XXX. Received YYY; in original form ZZZ}

\pubyear{2018}

\begin{document}
\label{firstpage}
\pagerange{\pageref{firstpage}--\pageref{lastpage}}
\maketitle

\begin{abstract}
We introduce a pixel space method to detect dipole modulation or hemispherical power asymmetry in the cosmic microwave background (CMB) polarization. The method relies on the use of squared total polarized flux whose ensemble average picks up a dipole due to the dipole modulation in the CMB polarization. The method is useful since it can be applied easily to partial sky. We define several statistics to characterize the amplitude of the detected signal. By simulations we show that the method can be used to reliably extract the signal at 2.7$\sigma$ level or higher in future CORE-like missions, assuming that the signal is present in the CMB polarization at the level detected by the Planck mission in the CMB temperature. An application of the method to the 2018 Planck data does not detect a significant effect, when taking into account the presence of correlated detector noise and residual systematics in data. Using the FFP10 we find the presence of a very strong bias which might be masking any real effect.
\end{abstract}




\section{Introduction}

The hemispherical power asymmetry (HPA) is one of the most well known and well-studied anomaly observed in the cosmic microwave background (CMB) temperature fluctuation field. It is an important observation questioning the validity of the Cosmological Principle. The Cosmological Principle is an assumption that the spatial distribution of matter and radiation in the universe is statistically isotropic. The HPA is the observation that with the z axis along ($l=232^\circ, b=-14^\circ$), the power of the CMB temperature is slightly higher in the northern hemisphere compared to the southern hemisphere. It was originally observed in the Wilkinson Microwave Anisotropy Probe (WMAP) data \citep{Eriksen:2003,Eriksen:2007,Hansen:2008,Hoftuft:2009,Paci:2013,Prunet:2004}. The phenomenon has persisted in the Planck CMB temperature fluctuation data \citep{Ade:2013,Akrami:2014,Ade:2015,Rath:2013,Ghosh:2016}. 

The HPA is not the only observation of statistical isotropy (SI) violation. Other observations of SI violations include the CMB temperature quadrupole-octopole alignment \citep{deOliveira-Costa:2004,Schwarz:2004,Ralston:2004,Samal:2008}, the octopole planarity \citep{Bennett:2010}, the CMB parity anomaly \citep{Kim:2010,Aluri:2012}, the CMB temperature and polarization multipole alignments \citep{Rath:2017,Pinkwart:2018}, the CMB Cold Spot \citep{Cruz:2004}, the kinematic dipole excess in the large-scale structures \citep{Singal:2011,Gibelyou:2012,Rubart:2013,Tiwari:2014,Tiwari:2013,Tiwari:2015,Bengaly:2017,Rameez:2017}, and the radio and optical polarization alignments \citep{Huts:1998,Jain:2006,Tiwari:2012} and radio dipole \citep{Jain:1998kf}. A review of the SI violating phenomena can be found in \citet{Ghosh:2017}.   

\citet{Gordon:2006} proposed a dipole modulation model to explain HPA in temperature. According to this model, the observed temperature fluctuation field $\Delta T (\boldsymbol{\hat n})$ can be written as:
\begin{equation}
    \Delta T (\boldsymbol{\hat n}) = \Delta \tilde T (\boldsymbol{\hat n})\left[ 1 + A \boldsymbol{\hat \lambda} \cdot \boldsymbol{\hat n} \right].
    \label{Eq:T_mod}
\end{equation}
Here $\Delta \tilde T (\boldsymbol{\hat n})$ is a statistically isotropic field which is modulated by a cosine modulation term $\boldsymbol{\hat \lambda} \cdot \boldsymbol{\hat n}$ with an amplitude of $A$ and direction $\boldsymbol{\hat \lambda}$. There are other theoretical models which have been proposed in an attempt to explain the HPA in CMB temperature fluctuations \citep{Groeneboom:2010,Rath:2013a,Rath:2014, Bridges:2006,Boehmer:2007,Carroll:2008,Erickcek:2008,Erickcek:2009,Emami:2010,Aluri:2011,Chang:2013,Cai:2013,Zibin:2015}. A dipole modulation model, Eq. \eqref{Eq:T_mod}, as well as other theoretical models predict that one should be able to observe dipole modulation phenomenon in CMB polarization too \citep{Namjoo:2014,Kothari:2015,Mukherjee:2015,Ghosh:2016,Contreras:2017}.

The estimators employed to search for the dipole modulation in the CMB temperature can be broadly classified as those methods working in multipole space with spherical harmonic coefficients of the CMB map \citep{Hansen:2008,Rath:2013,Zibin:2015,Hajian:2003} and those methods which are applied on the maps directly in pixel space \citep{Eriksen:2003,Rath:2013,Akrami:2014}. For detection of dipole modulation in CMB polarization there are correspondingly different methods that work in multipole space \citep{Basak:2006,Ghosh:2016,Contreras:2017} or pixel space \citep{Aluri:2017}. There are also maximum likelihood-based estimation techniques employed for detecting HPA in CMB temperature and polarization \citep{Gordon:2006,Hoftuft:2009}. In this work, we will introduce a procedure for estimating the dipole modulation amplitude and direction in pixel space using CMB polarized intensity maps.


\section{Theory} \label{Sec:Theory}
The signal of the CMB polarization is measured as Stokes parameters $Q(\boldsymbol{\hat n})$ and $U(\boldsymbol{\hat n})$. The quantity $P(\boldsymbol{\hat n}) = Q(\boldsymbol{\hat n}) + i U(\boldsymbol{\hat n})$ behaves as a spin-2 object while its complex conjugate $P^*(\boldsymbol{\hat n}) = Q(\boldsymbol{\hat n}) - i U(\boldsymbol{\hat n})$ behaves as a spin-($-2$) object. When measured in an experiment the signal is observed along with instrumental noise. We write the resulting signal as:
\begin{equation}
    \begin{aligned}
    P(\boldsymbol{\hat n}) &= P_s(\boldsymbol{\hat n}) + N_P(\boldsymbol{\hat n})\\
    P^*(\boldsymbol{\hat n}) &= P^*_s(\boldsymbol{\hat n}) + N^*_P(\boldsymbol{\hat n}).
\end{aligned}
\label{Eq:SigNoi}
\end{equation}
Here $P_s(\boldsymbol{\hat n}) = Q_s(\boldsymbol{\hat n}) + i U_s(\boldsymbol{\hat n})$ is the CMB polarization signal and $N_P(\boldsymbol{\hat n}) = N_Q(\boldsymbol{\hat n}) + i N_U(\boldsymbol{\hat n})$ is the combination instrumental noise in $Q$ and $U$ measurements. These quantities can be expanded in spin($\pm$2) spherical harmonics as:
\begin{align}
    P_{X}(\boldsymbol{\hat n}) &= \sum_{\ell,m}a^{X}_{2,\ell m} {}_2Y_{\ell m} (\boldsymbol{\hat n}) = -\sum_{\ell,m} (a^X_{E,\ell m} + i a^X_{B,\ell m}) {}_2Y_{\ell m} (\boldsymbol{\hat n}) \label{Eq:Pexp}\\
    P^*_{X}(\boldsymbol{\hat n}) &= \sum_{\ell,m}a^{X}_{-2,\ell m} {}_{-2}Y_{\ell m} (\boldsymbol{\hat n}) \nonumber\\ 
    &= -\sum_{\ell,m} (a^X_{E,\ell m} - i a^X_{B,\ell m}) {}_{-2}Y_{\ell m} (\boldsymbol{\hat n}) \nonumber \\
    & =-\sum_{\ell,m} (a^{X*}_{E,\ell m} - i a^{X*}_{B,\ell m}) {}_2Y^*_{\ell m} (\boldsymbol{\hat n}), \label{Eq:P*exp} 
\end{align}
with $X$ being index for observed, signal or noise components of Eq. \eqref{Eq:SigNoi}.
Here $a^X_{E, \ell m}$ and $a^X_{B, \ell m}$ are the spherical harmonic coefficients of E-mode and B-mode polarization. 

We will consider the dipole modulation in CMB polarization modelled as in \citet{Ghosh:2016}. This modulation has been motivated from a physical perspective in \citet{Contreras:2017}. The CMB polarization signal $P_s(\boldsymbol{\hat n})$ is modulated as:
\begin{equation}
    P_s(\boldsymbol{\hat n}) = \tilde P_s(\boldsymbol{\hat n}) \left(1 + A \boldsymbol{\hat \lambda} \cdot \boldsymbol{\hat n}\right).
    \label{Eq:ModRel}
\end{equation}
Here $\tilde P_s(\boldsymbol{\hat n})$ is the unmodulated isotropic polarization field, $A$ and $\boldsymbol{\hat \lambda}$ are the amplitude and direction of the dipole modulation respectively. The relation Eq. \eqref{Eq:ModRel} would imply that the modulation amplitude $A$ is in general a complex number, as assumed by \citet{Ghosh:2016}. However, modulation of the form in Eq. \eqref{Eq:ModRel} results from a modulation of the temperature anisotropy quadrupole at the scatterers \citep{Contreras:2017}. This implies that the modulation amplitude $A$ is real. From current observational evidence from CMB temperature anisotropies we can assume that the modulation amplitude is small and confine ourselves to first order in $A$. It is also well known that the hemispherical power asymmetry in CMB temperature is a scale dependent phenomenon. The effect is pronounced at large angular scales and disappears by $\ell\sim100$ \citep{Hansen:2008,Hanson:2009,Rath:2014,Aiola:2015,Quartin:2015}. This implies that the dipole modulation amplitude $A$ in Eq. \eqref{Eq:T_mod} and Eq. \eqref{Eq:ModRel} should be scale dependent. While the overall amplitude in the $2\le \ell \le 64$ range is well known, the exact form of the scale dependence of the amplitude is not. Hence we will assume the simple scale independent modulation model of Eq. \eqref{Eq:T_mod} with low resolution map to look at the implications for CMB polarization. The constant amplitude $A$ may be replaced by $A_0(\ell/\ell_0)^{-\alpha}$ as a simple extension to the minimal model. We will postpone it to a future work. We point out that it is not possible to introduce the dipole modulation directly into the $E$ mode polarization field \citep{Kothari:2018}. Furthermore the off-diagonal TE correlations are in general different from the corresponding ET correlations in the presence of dipole modulation \citep{Kothari:2015}.  

All the expressions written till now are for full sky observations. In reality, we only observe a fraction of the sky as galactic plane and point sources have to be masked during analysis. Let us assume a function $W(\boldsymbol{\hat n})$ as the mask function. We represent the observed masked sky CMB polarization as $P_{\text{obs}}(\boldsymbol{\hat n})=P(\boldsymbol{\hat n})W(\boldsymbol{\hat n})$. This is given by:
\begin{equation}
    P_{\text{obs}} = \tilde P_s(\boldsymbol{\hat n}) W(\boldsymbol{\hat n}) \left(1 + A \boldsymbol{\hat \lambda} \cdot \boldsymbol{\hat n}\right) + N_P(\boldsymbol{\hat n}) W(\boldsymbol{\hat n}).
\end{equation}

We are interested in the quantity $|P_{\text{obs}}(\boldsymbol{\hat n})|^2$ for a masked sky. We obtain 
\begin{align}
    |P_{\text{obs}} (\boldsymbol{\hat n})|^2 =& |\tilde P_s(\boldsymbol{\hat n})|^2 W^2(\boldsymbol{\hat n}) \left(1 + 2A \boldsymbol{\hat \lambda} \cdot \boldsymbol{\hat n}\right) \nonumber\\ 
    & + \tilde P_s(\boldsymbol{\hat n}) N^*_P(\boldsymbol{\hat n}) W^2(\boldsymbol{\hat n}) \left(1 + A \boldsymbol{\hat \lambda} \cdot \boldsymbol{\hat n}\right) \nonumber\\ 
    & + \tilde P^*_s(\boldsymbol{\hat n}) N_P(\boldsymbol{\hat n}) W^2(\boldsymbol{\hat n}) \left(1 + A \boldsymbol{\hat \lambda} \cdot \boldsymbol{\hat n}\right) \nonumber\\
    &+|N_P(\boldsymbol{\hat n})|^2 W^2(\boldsymbol{\hat n})
    \label{Eq:modP2}
\end{align}
We now substitute the expansion of these quantities from Eq. \eqref{Eq:Pexp} and Eq. \eqref{Eq:P*exp} and take realization average. 
For the unmodulated polarization field the realization average of the spherical harmonic coefficients satisfies:
\begin{equation}
    \begin{aligned}
        \langle \tilde a_{E,\ell m} \tilde a^*_{E,\ell' m'}\rangle &= C_\ell^{EE}\delta_{\ell \ell'} \delta_{m m'}\\
        \langle \tilde a_{B,\ell m} \tilde a^*_{B,\ell' m'}\rangle &= C_\ell^{BB}\delta_{\ell \ell'} \delta_{m m'}\\
        \langle \tilde a_{E,\ell m} \tilde a^*_{B,\ell' m'}\rangle &= \langle \tilde a_{B,\ell m} \tilde a^*_{E,\ell' m'}\rangle = 0,
    \end{aligned}
    \label{Eq:SigCls}
\end{equation}
In these expressions $C_\ell ^{EE}$ and $C_\ell^{BB}$ are the CMB E-mode and B-mode power spectrum. While the noise satisfies:
\begin{equation}
    \begin{aligned}
        \langle n_{E,\ell m} n^*_{E,\ell' m'}\rangle &= N_\ell^{EE}\delta_{\ell \ell'} \delta_{m m'}\\
        \langle n_{B,\ell m} n^*_{B,\ell' m'}\rangle &= N_\ell^{BB}\delta_{\ell \ell'} \delta_{m m'}\\
        \langle n_{E,\ell m} n^*_{B,\ell' m'}\rangle &= \langle n_{B,\ell m} n^*_{E,\ell' m'}\rangle = 0 \\
    \end{aligned}
    \label{Eq:NoiCls}
\end{equation}
along with $\langle \tilde a_{X,\ell m} n^*_{Y,\ell' m'}\rangle =\langle n_{X,\ell m} \tilde a^*_{Y,\ell' m'}\rangle = 0$, for $X,Y$ in $\{E,B\}$. The quantities $N_\ell ^{EE}$ and $N_\ell^{BB}$ are E-mode and B-mode noise power spectrum. Using relations Eq. \eqref{Eq:SigCls} and Eq. \eqref{Eq:NoiCls} we can write the realization average of $|P_{\text{obs}}(\boldsymbol{\hat n})|^2$ as:
\begin{align}
    \langle |P_{\text{obs}}(\boldsymbol{\hat n})|^2\rangle=& \sum_\ell \left\{ \left[C_\ell^{EE} + C_\ell^{BB}\right] \left[1+2A\boldsymbol{\hat \lambda} \cdot \boldsymbol{\hat n}\right] \right.\nonumber\\ &\left. +\left[N_\ell^{EE} + N_\ell^{BB}\right] \right\}  W^2(\boldsymbol{\hat n})  \sum_m {}_2Y_{\ell m}(\boldsymbol{\hat n}) {}_2Y^*_{\ell m}(\boldsymbol{\hat n}).
    \label{Eq:P2ClRel}
\end{align}
Using the generalized addition theorem for Wigner D-functions
\citep{Varshalovich:1988}, we obtain,
\begin{equation}
    \sum_m{}_{2}Y_{\ell m}(\boldsymbol{\hat n}) {}_{2}Y^*_{\ell m} (\boldsymbol{\hat n}) = {\frac{(2\ell+1)}{4\pi}} ,
\end{equation}
 Substituting back in our original expression Eq. \eqref{Eq:P2ClRel} we get:
\begin{align}
    \langle |P_{\text{obs}}(\boldsymbol{\hat n}_i)|^2\rangle =&  W^2(\boldsymbol{\hat n}_i) \sum_\ell \left(\frac{2\ell + 1}{4\pi}\right) \left\{\left[\bar C_\ell^{EE} + \bar C_\ell^{BB}\right] \right. \nonumber\\ & \left. \times \left[1+2 A\boldsymbol{\hat \lambda} \cdot \boldsymbol{\hat n}_i\right] + \left[\bar N_\ell^{EE} + \bar N_\ell^{BB}\right] \right\}.
    \label{Eq:P2maskedfin}
\end{align}
For realistic analysis we have replaced $\boldsymbol{\hat n} $ and $\theta$ with $\boldsymbol{\hat n}_i$ and $\theta_i$ of the $i^\text{th}$ pixel and the barred power spectra represent adjustment for appropriate beam function and pixel window viz $\bar C^{XX}_\ell = C^{XX}_\ell B_\ell^2 F_\ell^2$. Note that $B_\ell$ is the beam function and $F_\ell$ is the pixel window function at the resolution at which we will perform the analysis.


\subsection{Estimator construction} \label{Sec:estimator}
It is evident from Eq. \eqref{Eq:P2maskedfin} that $\langle |P|^2\rangle$ will vary as a dipole due to the $\boldsymbol{\hat \lambda} \cdot \boldsymbol{\hat n}$ term. We define a cosine weighted averaged quantities as:
\begin{align}
    \langle |P_\text{obs}(\boldsymbol{\hat n})|^2 \rangle_w = \frac{\int_{\Omega_r} |P_\text{obs}(\boldsymbol{\hat n})|^2  \cos \theta d\Omega}{\int_{\Omega_r}W(\boldsymbol{\hat n}) \cos \theta d\Omega} \equiv \frac{\int_{\Omega_r} |P_\text{obs}(\boldsymbol{\hat n})|^2  \cos \theta d\Omega}{\int_{\Omega_r} W^2(\boldsymbol{\hat n})\cos \theta d\Omega}
    \label{Eq:cosWP2def}
\end{align}
where the last equivalence holds for a binary mask satisfying $W(\boldsymbol{\hat n}) = W^2(\boldsymbol{\hat n})$. Here $\theta$ is the angle that direction $\boldsymbol{\hat n}$ subtends from the z axis.  This equivalence would not hold for apodized masks. However in pixel space analysis we do not use apodization, justifying the approximation made in the equivalence. Note that, since the cosine weight multiplied here is non-stochastic, a cosine weighted stochastic average of $|P_\text{obs}(\boldsymbol{\hat n})|^2$ is equivalent to multiplying both sides of Eq \eqref{Eq:P2maskedfin} with a $\cos \theta$ factor.

We are interested in fitting the dipole in $|P_\text{obs}(\boldsymbol{\hat n})|^2$ for our parameters $\boldsymbol{\hat \lambda}$ and $A$. For this fitting, we choose an $i^\text{th}$ pixel to be the $\boldsymbol{\hat z}$ direction and divide the spherical sky into two hemispheres with the $\boldsymbol{\hat z}$ pointing outward through the center of the upper hemisphere and the $-\boldsymbol{\hat z}$ pointing outward through the center of the lower hemisphere. For every $j^\text{th}$ pixel on the sphere we define a weight function $w_{i,j}=|\cos \theta_{i,j}|$, where $\theta_{i,j}$ is the co-latitude of the $j^\text{th}$ pixel for the choice of the $i^\text{th}$ pixel as the z axis. On each of the hemisphere, we can find the average value of our $|P|^2$ field weighted by the $w_{i,j}$ weight factor as: 
\begin{equation}
    \langle |P_\text{obs}(\boldsymbol{\hat n})|^2\rangle_w = \frac{\sum_j w_{i,j}|P_\text{obs}(\boldsymbol{\hat n}_j)|^2}{\sum_{j'}w_{i,j'}W(\boldsymbol{\hat n}_{j'})}
    \label{Eq:weightedP2}
\end{equation}
By weighting the pixels of the maps we are effectively searching for a cosine (dipolar) variation in the $|P|^2$ field. Then, for every choice of $\boldsymbol{\hat z}$ along $i^\text{th}$ pixel, we can define two statistics $R_i$ and $D_i$ as follows:
\begin{align}
    R_i &= \frac{\langle |P_\text{obs}(\boldsymbol{\hat n})|^2\rangle_{w,U_i}}{\langle 
|P_\text{obs}(\boldsymbol{\hat n})|^2\rangle_{w,D_i}} \label{Eq:Rstat} \\
    D_i &= \frac{\langle |P_\text{obs}(\boldsymbol{\hat n})|^2\rangle_{w,U_i} - 
    \langle |P_\text{obs}(\boldsymbol{\hat n})|^2\rangle_{w,D_i}}{\langle |P_\text{obs}(\boldsymbol{\hat n})|^2\rangle_{w,U_i} + 
    \langle |P_\text{obs}(\boldsymbol{\hat n})|^2\rangle_{w,D_i}} \label{Eq:Dstat}.
\end{align}
Here $\langle |P_\text{obs}(\boldsymbol{\hat n})|^2\rangle_{w,U_i}$ and $\langle |P_\text{obs}(\boldsymbol{\hat n})|^2\rangle_{w,D_i}$ denote the weighted average value of the $|P_\text{obs}|^2$ 
field on the upper and lower hemispheres respectively for $\boldsymbol{\hat z}$ along $\boldsymbol{\hat n}_i$. We maximize these statistics by making a search over $\boldsymbol{\hat z}$. This 
resulting direction is the preferred axis $\boldsymbol{\hat \lambda}$.  
Assuming that the dipole modulation effect is small both 
statistics will lead to the similar result. 
The $|P|^2$ map is supposed to contain a dipole due to the dipole modulation. Hence,  we may also directly extract the power spectrum $C_\ell$ as well as the spherical harmonic coefficients for $\ell=1$ from the masked sky by spherical harmonic transformation and suitably accounting for the masking. The dipole power ($C_1$) provides us with another estimate of the dipole modulation effect in data as $\sqrt{C_1} \propto A$.

\begin{table}
    \centering
    \begin{tabular}{lccc}
    \hline
        $\nu$ & $\theta_\text{FWHM}$ & $\Delta T$ & $\Delta P$ \\
        GHz & [arcmin] & [$\mu$K arcmin]& [$\mu$K arcmin]\\
    \hline
        60 &  17.87  &  7.5   &  10.6\\
        70 &  15.39  &  7.1   &  10.0\\
        80 &  13.52  &  6.8   &  9.6\\
        90 &  12.08  &  5.1   &  7.3\\
        100 &  10.92  &  5.0   &  7.1\\
        115 &  9.56   &  5.0   &  7.0\\
        130 &  8.51   &  3.9   &  5.5\\
        145 &  7.68   &  3.6   &  5.1\\
        160 &  7.01   &  3.7   &  5.2\\
        175 &  6.45   &  3.6   &  5.1\\
        195 &  5.84   &  3.5   &  4.9\\
        220 &  5.23   &  3.8   &  5.4\\
        255 &  4.57   &  5.6   &  7.9\\
        295 &  3.99   &  7.4   &  10.5\\
        340 &  3.49   &  11.1  &  15.7\\
        390 &  3.06   &  22.0  &  31.1\\
        450 &  2.65   &  45.9  &  64.9\\
        520 &  2.29   &  116.6 &  164.8\\
        600 &  1.98   &  358.3 &  506.7\\
    \hline
    \end{tabular}
    \caption{CORE 4 year mission performance summary taken from \citet{Delabrouille:2018}. Here $\Delta T$ and $\Delta P$ is the white noise level in $\mu K$ in an arcmin size pixel in temperature and polarization respectively.}
    \label{tab:COREnoise}
\end{table}

Once we have identified the direction along which either $R_i$ or $D_i$ estimators maximise we have our best estimate for $\boldsymbol{\hat\lambda}$. First we assume full sky data with no sky masking. Using Eqs. \eqref{Eq:P2maskedfin} and \eqref{Eq:cosWP2def} we can see that the upper and lower hemisphere difference between the cosine weighted average value of $|P|^2$ is given by:
\begin{align}
    \left[\langle |P_\text{obs}(\boldsymbol{\hat n})|^2\rangle_{w,U} - 
\langle |P_\text{obs}(\boldsymbol{\hat n})|^2\rangle_{w,D} \right]_\text{max} =& \sum_\ell \left(\frac{2\ell + 1}{4\pi}\right) \nonumber \\ & \left(\bar C_\ell^{EE} + \bar C_\ell^{BB}\right) \frac{8A}{3}.
\end{align}
We have set $\boldsymbol{\hat z}$ along $\boldsymbol{\hat \lambda}$. Hence we can write the estimator for the amplitude of modulation as:
\begin{equation}
    \hat A = \frac{\left[\langle |P_\text{obs}(\boldsymbol{\hat n})|^2\rangle_{w,U} - 
\langle |P_\text{obs}(\boldsymbol{\hat n})|^2\rangle_{w,D} \right]_\text{max}}{\frac{4}{3}\sum_\ell \left(\frac{2\ell + 1}{2\pi}\right) \left(\bar C_\ell^{EE} + \bar C_\ell^{BB}\right)}
\label{Eq:AestFS}
\end{equation}
For a partial sky with binary mask $W(\boldsymbol{\hat n}_i)$ the $\hat A$ estimator changes as:
\begin{equation}
    \hat A = \frac{\left[\langle |P_\text{obs}(\boldsymbol{\hat n})|^2\rangle_{w,U} - 
\langle |P_\text{obs}(\boldsymbol{\hat n})|^2\rangle_{w,D} \right]_\text{max}}{K\sum_\ell \left(\frac{2\ell + 1}{2\pi}\right) \left(\bar C_\ell^{EE} + \bar C_\ell^{BB}\right)},
\label{Eq:AestPS}
\end{equation}
where $K$ is given by:
\begin{align}
    K = &\left\{\int_U W^2(\boldsymbol{\hat n})\cos \theta d\Omega \right\}^{-1} \int_U W^2(\boldsymbol{\hat n}) \cos^2 \theta d\Omega \nonumber \\ &- \left\{\int_D W^2(\boldsymbol{\hat n}) \cos \theta d\Omega \right\}^{-1} \int_D W^2(\boldsymbol{\hat n}) \cos^2 \theta d\Omega.
    \label{Eq:Krel}
\end{align}
We maximize the $R$ and $D$ statistics to find the preferred direction 
$\boldsymbol{\hat \lambda}$. Along the direction $\boldsymbol{\hat \lambda}$ we use the $\hat A$ estimator to find the amplitude. We have tested the estimators with isotropic and modulated simulations and used these with Planck 2018 polarization data.

\section{Simulation} \label{Sec:Sim}
We used 2018 Planck cosmological parameters \citep{Aghanim:2018} to generate the lensed power spectrum for scalar perturbations using CAMB\footnote{http://camb.info}. This power spectrum is used with \texttt{synfast} facility from HEALPix\footnote{http://healpix.sourceforge.net} to generate isotropic CMB sky maps for our forecasts. For our forecasts, we used Gaussian beams with FWHM equal to three times the pixel size. For the 2018 Planck legacy data we used the Full Focal Plane 10 (FFP10) simulations made available through the Planck Legacy Archive\footnote{http://pla.esac.esa.int/pla/}. We used the first 300 FFP10 CMB simulation with their corresponding noise-and-systematics simulations. The FFP10 CMB maps and the noise-and-systematics maps were used for the significance testing of our results for the 2018 Planck data. The FFP10 CMB files for different component separation method are named: ``\texttt{dx12\textunderscore v3\textunderscore<COM\textunderscore SEP>\textunderscore cmb\textunderscore mc\textunderscore00XXX.fits}''. where \texttt{<COM\textunderscore SEP>} stands for a component separation method from any of Commander, SMICA, SEVEM and NILC. The \texttt{XXX} stands for the simulation number between 000 and 299.

\begin{figure}
    \centering
    \includegraphics[width=0.95\columnwidth]{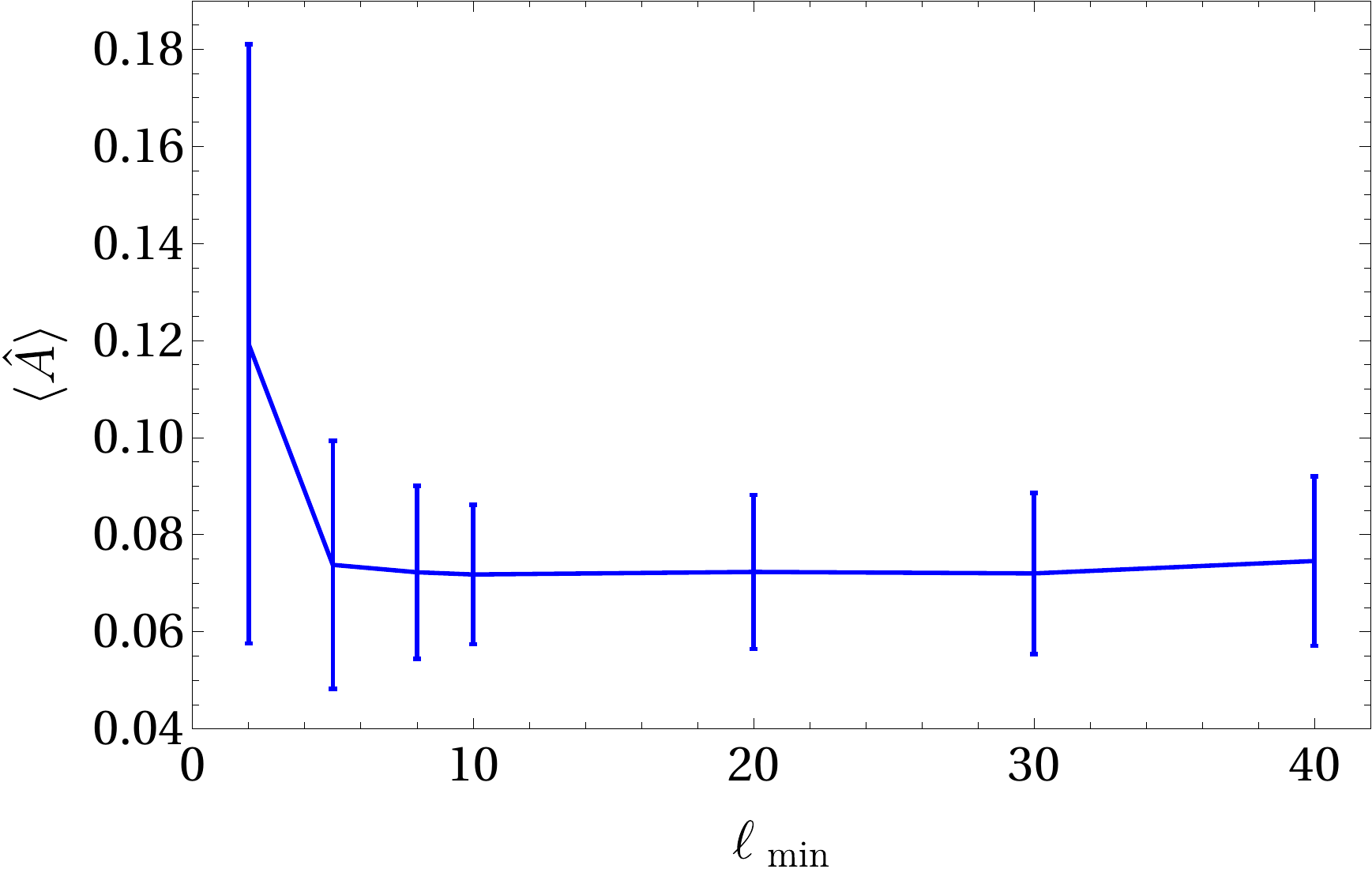}
    \caption{The plot shows the variation of the mean of the estimator $\hat A$ with $\ell_\text{min}$ using 150 simulated maps with input dipole modulation parameter $A=0.07$. Here $\ell_\text{min}$ is the multipole below which all modes are filtered out.}
    \label{Fig:A_v_lmin}
\end{figure}

\begin{figure}
    \centering
    \includegraphics[width=\columnwidth]{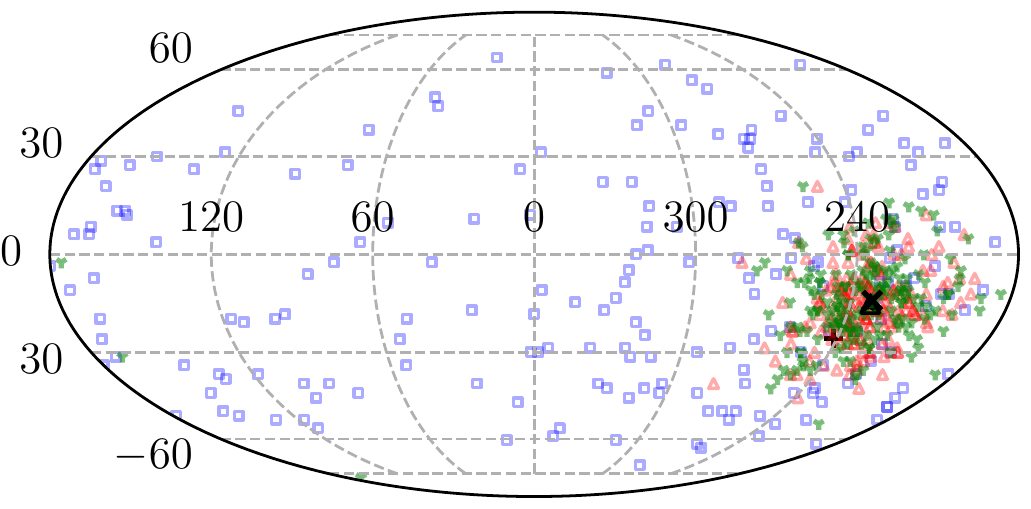}
    \caption{The scatter plot of the extracted directions using the $R$ estimator. The blue $\square$ markers are for $\ell_\text{min}=2$, red $\vartriangle$ markers are for $\ell_\text{min}=10$ and green $\Ydown$ markers are for $\ell_\text{min}=20$. The actual input direction is $l=232^\circ$, $b=-14^\circ$, denoted by black `$\times$' and the black `$+$', `$\Yup$' and `$\vartriangle$' denote the mean of the estimated values with $\ell_\text{min}$ of 2, 10 and 20 respectively. Note that the mean of the estimated directions for $\ell_\text{min}$ of 10 and 20 coincide.}
    \label{Fig:lamb_lmin_var}
\end{figure} 

\subsection{Modulated Map Simulation}\label{Sec:SimMod}
For our forecasts we modulated the maps along the hemispherical power asymmetry direction in temperature fluctuations, $(l=232^\circ, b=-14^\circ)$. We assume a modulation amplitude of $A=0.07$ as observed in the CMB temperature data. For simplicity we assume $\boldsymbol{\hat \lambda}$ along $\boldsymbol{\hat z}$. Then $\boldsymbol{\hat \lambda} \cdot \boldsymbol{\hat n} = \sqrt{4\pi /3}Y_{10}(\boldsymbol{\hat n})$. We can expand both  sides of Eq. \eqref{Eq:ModRel} in terms of the relations in Eq. \eqref{Eq:Pexp} and Eq. \eqref{Eq:P*exp} and write  
\begin{align}
    a_{\pm 2,\ell m} = &\tilde a_{\pm 2,\ell m} + A\sqrt{\frac{4\pi}{3}} \sum_{\ell' m'}  \tilde a_{\pm 2,\ell' m'} \nonumber \\ & \times \int {}_{\pm 2}Y_{\ell' m'}(\boldsymbol{\hat n})Y_{1 0}(\boldsymbol{\hat n}){}_{\pm 2}Y^*_{\ell m}(\boldsymbol{\hat n}) d\Omega.
    \label{Eq:ModExp}
\end{align}
We can write the integral of Eq. \eqref{Eq:ModExp} in terms of Wigner-3j symbols using Gaunt's formula. Then rewriting $a_{\pm 2, \ell m}$s in terms of $a_{E,\ell m}$ and $a_{B, \ell m}$ and using relations Eq. \eqref{Eq:Pexp} and Eq. \eqref{Eq:P*exp} we get:
\begin{align}
    a_{E,\ell m} &= \tilde a_{E,\ell m} + A \alpha_- \tilde a_{E,\ell-1 m} + i A \alpha_0 \tilde a_{B,\ell m} + A \alpha_+ \tilde a_{E,\ell+1 m} \label{Eq:EalmMod}\\
    a_{B,\ell m} &= \tilde a_{B,\ell m} + A \alpha_- \tilde a_{B,\ell-1 m} - i A \alpha_0 \tilde a_{E,\ell m} + A \alpha_+ \tilde a_{B,\ell+1 m}, \label{Eq:BalmMod}
\end{align}
where,
\begin{align}
    & \alpha_- = \frac{1}{\ell}\sqrt{\frac{(\ell-2)(\ell+2)(\ell-m)(\ell+m)}{(2\ell-1)(2\ell+1)}} \\
    & \alpha_0 = \frac{2m}{\ell(\ell+1)}\\
    & \alpha_+ = \frac{1}{\ell+1}\sqrt{\frac{(\ell-1)(\ell+3)(\ell-m+1)(\ell+m+1)}{(2\ell+1)(2\ell+3)}}.
\end{align}
These relations are identical to those in \citet{Contreras:2017}. During implementation, the isotropic maps are generated by \texttt{synfast} at resolution of \texttt{NSIDE}=1024. These maps are used to generate the isotropic $\tilde a_{X, \ell m}$s. These are appropriately corrected for the beam function used in map generation and the pixel window function. These $\tilde a_{X,\ell m}$s are then modulated following \eqref{Eq:EalmMod} and \eqref{Eq:BalmMod} which are for $\boldsymbol{\hat \lambda}=\boldsymbol{\hat z}$. We preform appropriate rotations on the modulated $a_{X, \ell m}$s to rotate the $\boldsymbol{\hat \lambda}$ to the direction of modulation observed in CMB temperature: ($l=232^\circ,b=-14^\circ$). The modulated $a_{X, \ell m}$s are readjusted with suitable pixel window function and beam function. The modulated map is synthesized from $a_{X, \ell m}$s using \texttt{alm2map} utility.

\subsection{Noise Simulation} \label{Sec:SimNoise}
We used Planck FFP10 noise-and-systematics simulations for our Planck 2018 polarization data analysis. The 300 noise simulations include systematics simulations. The FFP10 noise files for different component separation are named as: ``\texttt{dx12\textunderscore v3\textunderscore<COM\textunderscore SEP>\textunderscore noise\textunderscore mc\textunderscore00XXX.fits}''.

When simulating noise for a CORE-like mission, we use the sensitivity and design specifications from \citet{Delabrouille:2018}. We reproduce this data in table \ref{tab:COREnoise}. With this data we generate the noise power spectrum using the following relation for instrumental noise as given by \citet{Errard-2015}:
\begin{equation}
    N_\ell^{XX} = \left[\sum_\nu w_{X,\nu}\exp{\left(-\ell(\ell+1)\frac{\theta_{\text{FWHM},\nu}^2}{8 \ln 2}\right)} \right]^{-1}.
\end{equation}
Here $w_{T,\nu} = (\Delta T)^{-2}$ and $w_{E/B,\nu} = (\Delta P)^{-2}$. We used the values of $\Delta T$ or $\Delta P$, the sensitivity and $\theta_\text{FWHM}$, the full width at half maximum in arc-minutes from table \ref{tab:COREnoise}. Following \citet{Errard-2015} we assume that there will be noise degradation due to component separation resulting in rescaling of $N_\ell^{XX}$ by a factor $\Delta$ which we assumed to be 1.5 for the present work. Hence the instrumental noise levels used in simulation was $\Delta \times N_\ell^{XX}$. We used \texttt{synfast} facility to generate random noise simulations from the noise power spectra. The noise maps are added to the simulated isotropic/modulated maps.

\begin{figure*}
    \centering
    \subfloat{\includegraphics[width=0.45\textwidth]{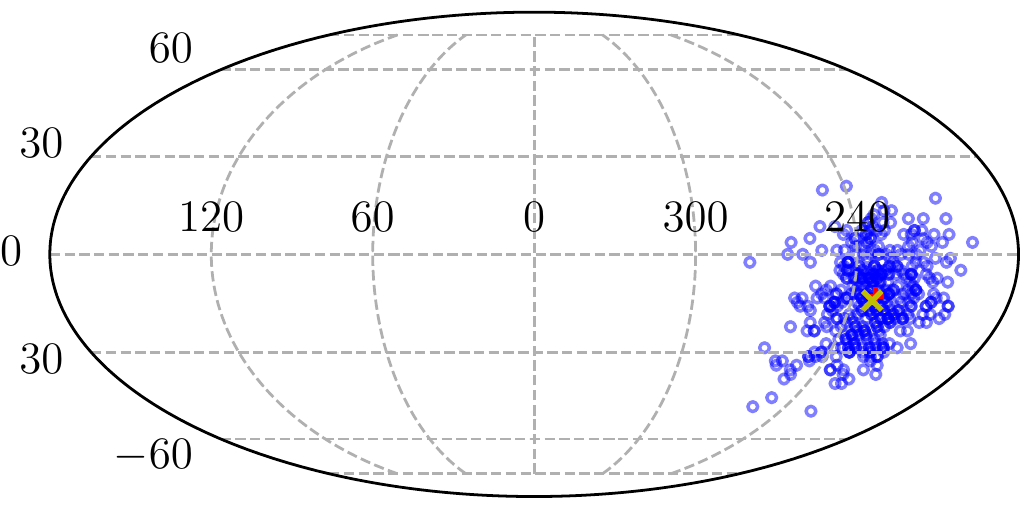}}\quad
    \subfloat{\includegraphics[width=0.45\textwidth]{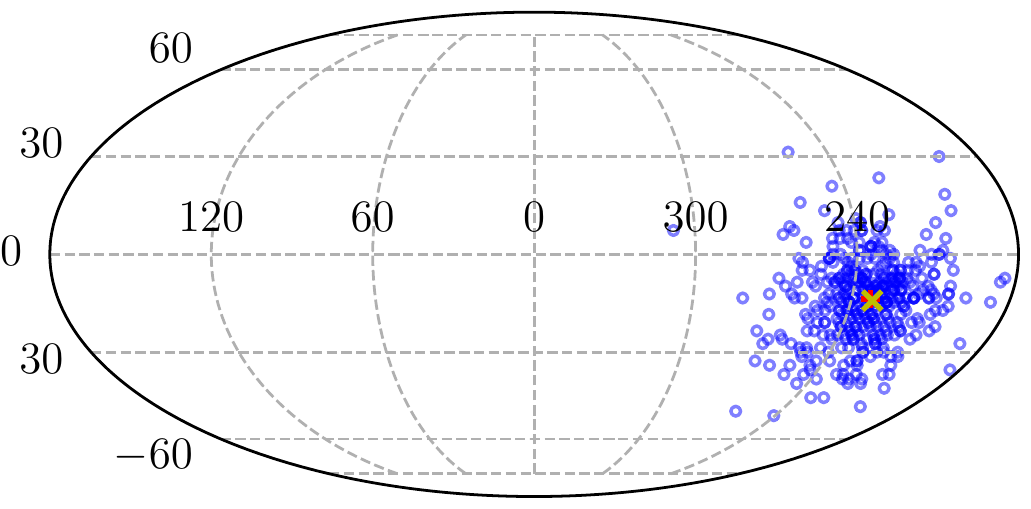}}\\
    \subfloat{\includegraphics[width=0.47\textwidth]{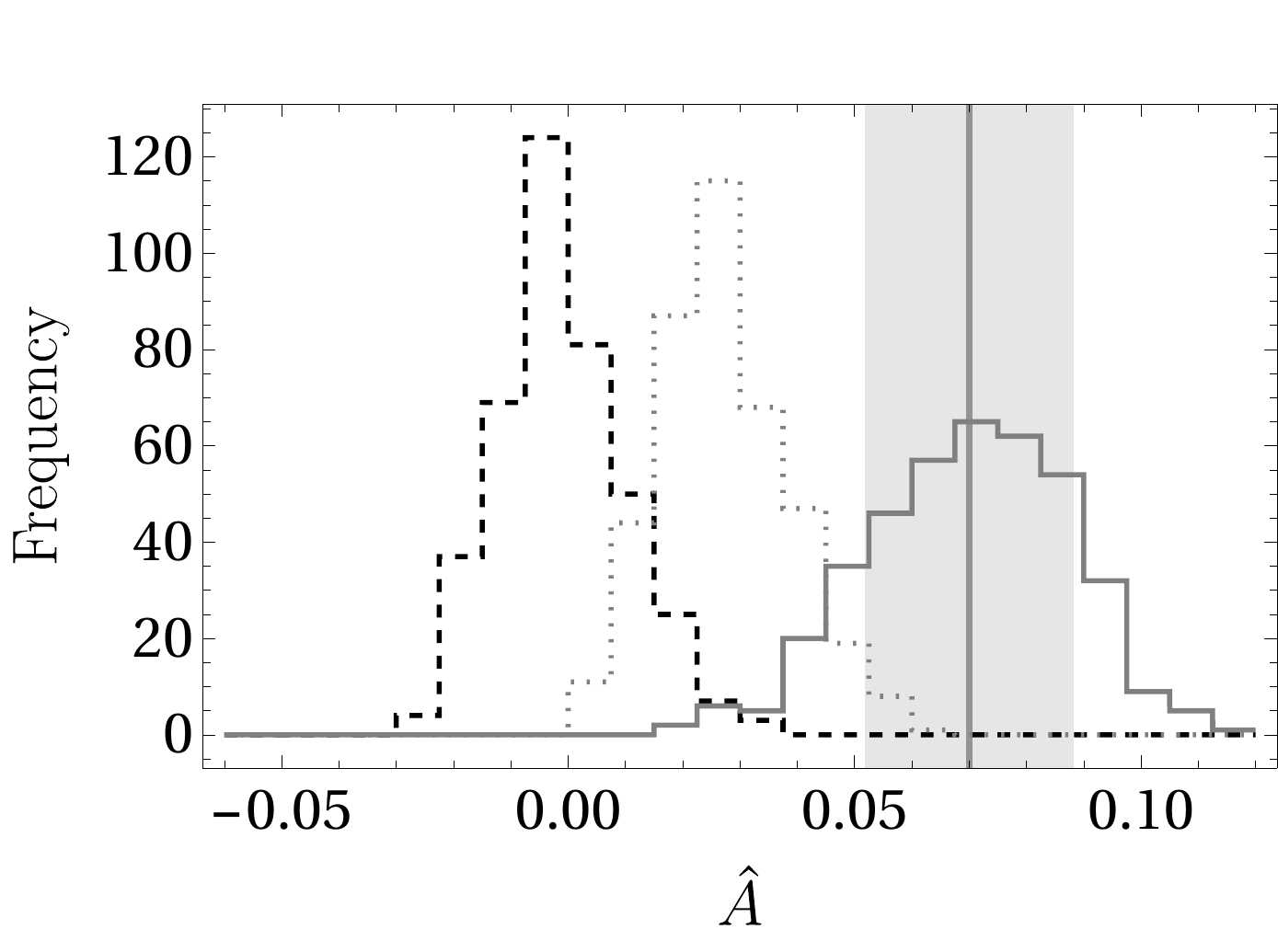}}\quad
    \subfloat{\includegraphics[width=0.47\textwidth]{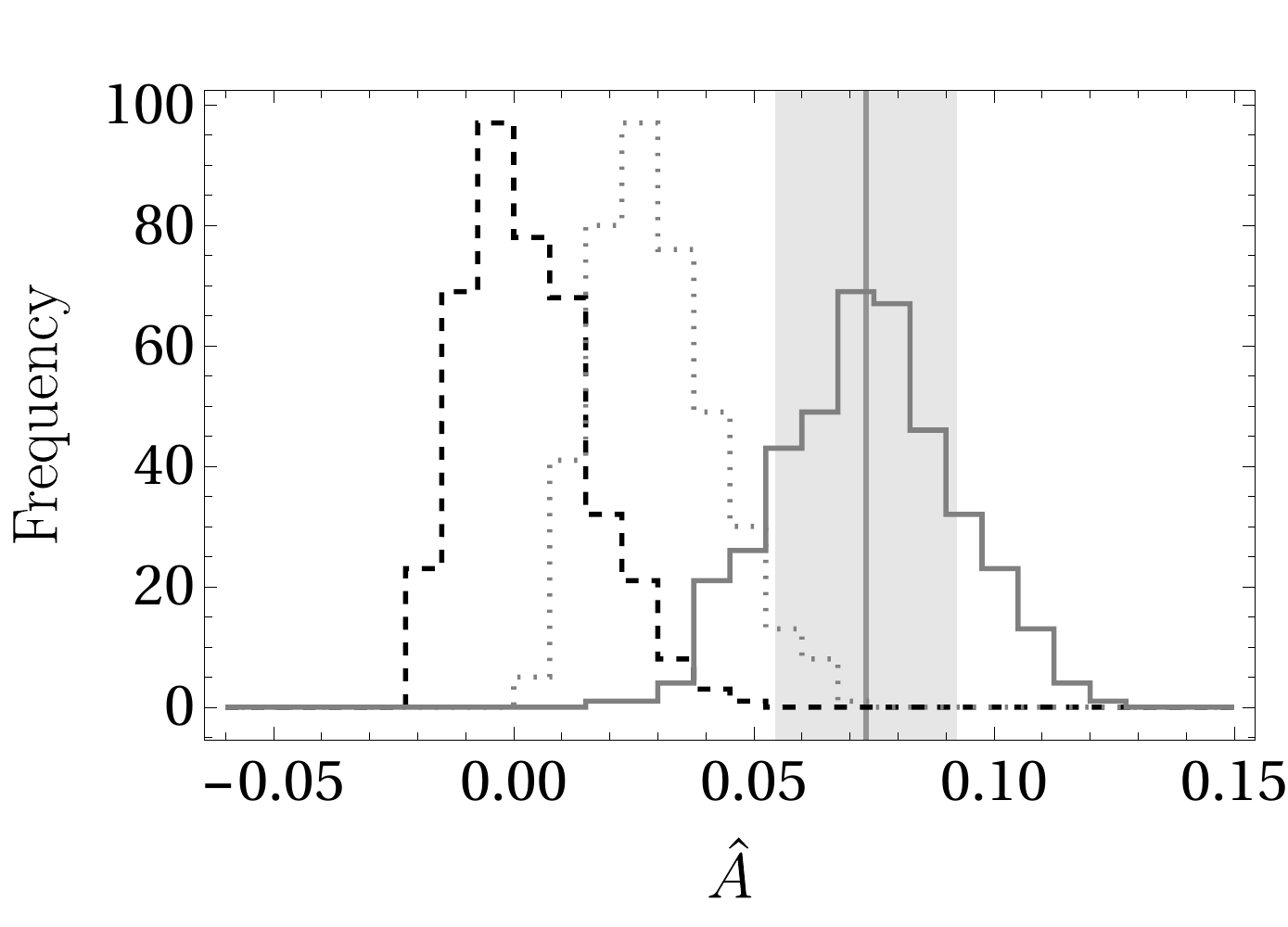}}\\
    \caption{The forecast for CORE-like mission simulation is shown in this figure. The left- column is for $\ell_\text{min}=10$ and the right column is $\ell_\text{min}=40$. The plots are for directions obtained by maximising the $R$ statistic. The results using the $D$ statistic are similar. The top row shows the direction scatter plots for 400 simulations shown with blue `$\circ$', The input direction of ($l=232^\circ, b=-14^\circ$) for the simulation is shown by the yellow `$\times$' and the red `$+$' indicates the mean of the estimated directions. The bottom row shows the histogram plots for the CORE-like simulations. The distribution of the estimated amplitude is shown for the 7$\%$ modulated distribution in solid grey, for the bias-corrected isotropic/null distribution in dashed, black and for the bias-uncorrected isotropic/null distribution is shown in dotted, grey. The mean estimated amplitude of the modulated distribution is shown by the black line, with the grey band indicating the 1-$\sigma$ range of the estimation. }
    \label{Fig:COREforecast}
\end{figure*}

\section{Data} \label{Sec:Data}
We use the 2018 Planck legacy polarization data (release 3.0) for this work. In the 2015 Planck data release the polarization data was high-pass filtered to remove the low-$\ell$ modes which contained systematic residuals \citep{Adam:2015}. In the 2018 data however all modes are present and the systematics have since been incorporated into the FFP10 noise simulations  \citep{Akrami:2018I,Akrami:2018II}. The 2018 Planck full mission legacy data files are named as: ``\texttt{COM\textunderscore CMB\textunderscore IQU-<COM\textunderscore SEP>\textunderscore2048\textunderscore R3.00\textunderscore full.fits}''.

The Planck polarization maps are provided at HEALPix \texttt{NSIDE}=2048. We downgrade our maps to \texttt{NSIDE} of 64 for the present analysis. We use the following relation for downgrading our maps:
\begin{equation}
    a_{\ell m}^\text{OUT} = \frac{a_{\ell m}^\text{IN}B^\text{OUT}_\ell F^\text{OUT}_\ell}{B^\text{IN}_\ell F^\text{IN}_\ell}.
    \label{Eq:DwnGrd}
\end{equation}
Here $B_\ell^\text{IN}$ correspond to the effective instrumental beam at native resolution as provided by Planck and $B_\ell^\text{OUT}$ correspond to Gaussian beam for the output resolution. The functions $F_\ell$ correspond to the pixel window functions at input and output resolutions. Using Eq. \eqref{Eq:DwnGrd} we prepare $a_{\ell m}$s between $\ell_\text{min}\leq \ell \leq \ell_\text{max}$, with $\ell_\text{max}=2\texttt{NSIDE}$. We will discuss the choice of \texttt{NSIDE} and $\ell_\text{min}$ in the next section. We use HEALPix subroutine \texttt{alm2map} to generate downgraded Q and U maps from these $a_{\ell m}$s. We prepare $|P|^2$ maps by calculating $Q^2+U^2$ in every pixel. For this entire work we will use the 2018 Planck common polarization mask: ``\texttt{COM\textunderscore Mask\textunderscore CMB-common-Mask-Pol\textunderscore 2048\textunderscore R3.00.fits}''. We will call this mask P18COM. The mask is not apodized. The mask is suitably downgraded and smoothed to \texttt{NSIDE} of  64.


\section{Method} \label{Sec:Method}
In section \ref{Sec:estimator} we defined our direction and amplitude estimators for dipole modulation. We discussed in section \ref{Sec:Theory} that the dipole modulation signal observed in CMB temperature is predominantly in the multipole range $2\le \ell \le 60$ and almost disappears by $\ell \sim 100$. The CMB transfer functions map the primordial perturbations in $k$-space to the $\ell$-space CMB temperature fluctuations and  CMB polarization observed today. However, the temperature and polarization transfer functions are different from each other. This implies that the $\ell$-space mapping of same physics for CMB polarization would be different from what would be observed CMB temperature. To make a prediction of a particular range we would have to assume a specific physical model beyond the minimal model assumed in Eq. \eqref{Eq:T_mod}. For model specific predictions one may refer to \citet{Kothari:2015} or \citet{Contreras:2017}. However, in absence of a physical model it is difficult to predict a particular $\ell$-range to look for the signal in polarization. It would still be reasonable to search for the effect in a similar multipole range in the CMB polarization data. So, we will analyse the data at a \texttt{NSIDE} of 64 with $\ell$ modes upto 128. This gives a $\ell_\text{max}$ cut of 128 for this work. For this purpose Q and U maps are degraded using Eq. \eqref{Eq:DwnGrd} and we produce $P^2$ maps by calculating $Q^2+U^2$ pixel by pixel. For all our analysis on data or on simulated maps we have used the P18COM mask. As described in section \ref{Sec:estimator}, we calculate $R_i$ and $D_i$ for every choice of $\boldsymbol{\hat z}$ on a \texttt{NSIDE}=64 pixelized sphere. In our analysis, we remove all multipoles below the $\ell_\text{min}$ cut, when preparing the maps. We find the direction along which $R$ or $D$ maximizes. This gives us the direction $\boldsymbol{\hat \lambda}$ of modulation. Along the direction of maximum $R$ and maximum $D$, we calculate the amplitude with use of Eq. \eqref{Eq:AestPS}. Note that the $C_\ell^{EE}$ and $C_\ell^{BB}$ used in Eq. \eqref{Eq:AestFS} or Eq. \eqref{Eq:AestPS} are multiplied with appropriate beam and pixel window functions.

\begin{figure}
    \centering
	\includegraphics[width=0.47\textwidth]{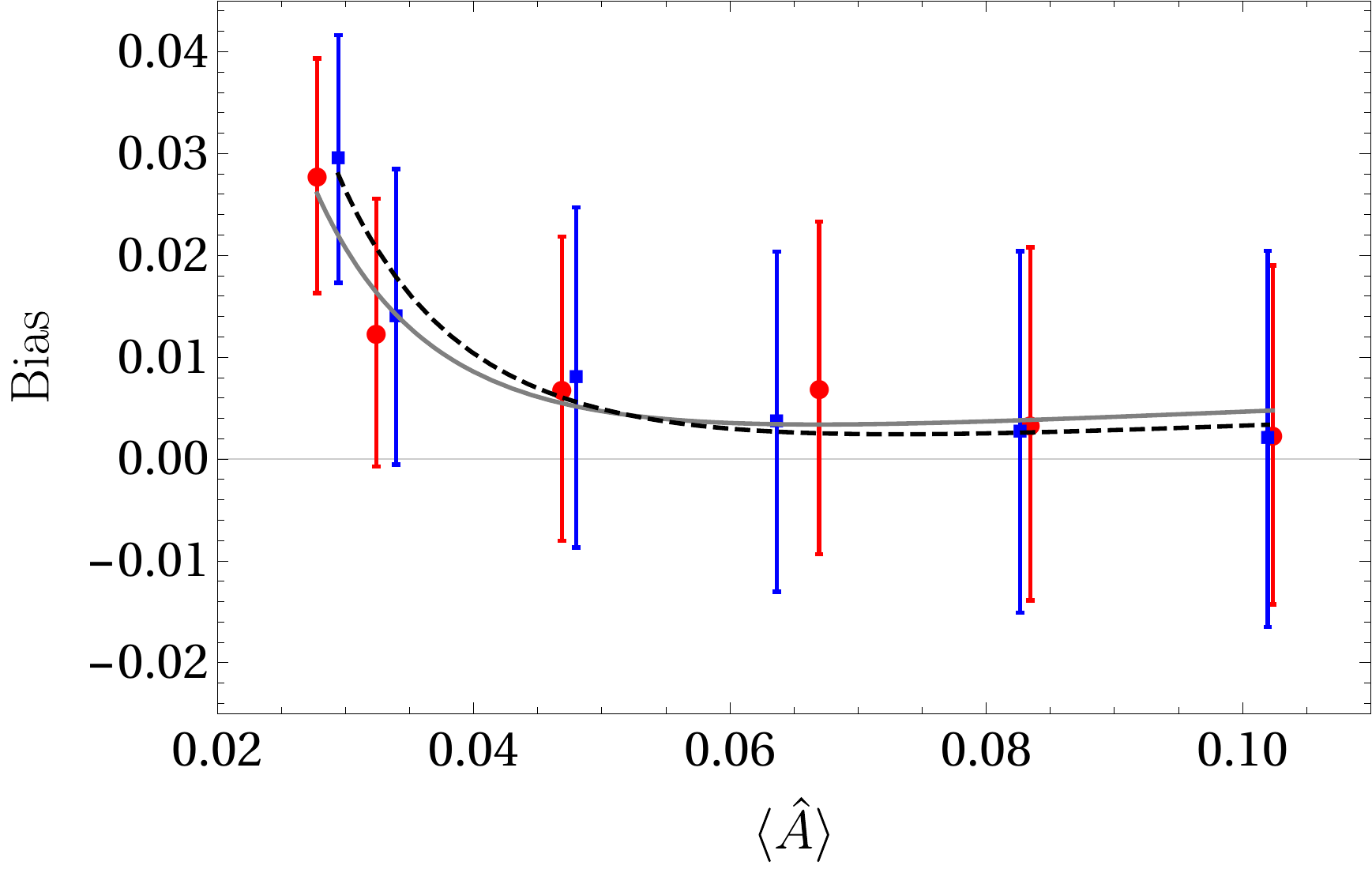}
    \caption{The variation of bias with $\hat A$ mean is shown in this plot. The red `$\bullet$' represent the simulations with $\ell_\text{min}=10$ and the blue `$\blacksquare$' are points with $\ell_\text{min}=40$.  The solid grey curve indicates the best-fit parametric model for bias with $\ell_\text{min}=10$, while the dashed, black curve is the parametric best-fit model for $\ell_\text{min}=40$. }
    \label{Fig:bias_var}
\end{figure}

\subsection{Low-$\ell$ cut}
The large angular scales, corresponding to the low-$\ell$ modes, have large uncertainties due to cosmic variance. Due to their large power contribution to the total sky power in large sky patches, such as a hemisphere, they significantly bias the mean value of the $P^2$ field. Since the low-$\ell$ modes show large 
fluctuation, their inclusion leads to a large bias in our estimators. 
In order to study this effect we generate 150 simulated maps with $7\%$ modulation and CORE-like noise. The extracted dipole amplitude from these simulations 
is shown in figure \ref{Fig:A_v_lmin} as a function of $ \ell_\text{min}$. Here $ \ell_\text{min}$ represents the cutoff value such that
for $\ell < \ell_\text{min}$ 
 we set all $a_{E,\ell m}$ and $a_{B,\ell m}$ to zero.
After this filtering of the low-$\ell$ modes, we resynthesize Q and U maps for use in our analysis. 
We find that with no filtering the output of $\hat A$ shows a large mean value with large uncertainties due to the fluctuations. We can see form figure \ref{Fig:A_v_lmin} that with variation in $\ell_\text{min}$ from 2 to 10 the mean of $\hat A$ decreases to the input value of 0.07 while the variance also decreases. From figure \ref{Fig:A_v_lmin} we can also see that with the variation of $\ell_\text{min}$ from 10 to 40, the mean of $\hat A$ remains largely unchanged and furthermore, the variance for the estimator increases very slightly. So the choice $\ell_\text{min}=10$ sufficiently removes the effects of cosmic variance. In figure \ref{Fig:lamb_lmin_var} we show the scatter plot  of directions extracted from the $R$ estimator for different values of $ \ell_\text{min}$. We find that for no filtering, i.e. $ \ell_\text{min}=2$, (shown with blue `$\square$') the directions  show considerable scatter all over the sky. With $\ell_\text{min}=10$, (denoted by red `$\vartriangle$') and $\ell_\text{min}=20$, (denoted by green `$\Ydown$') we see bias-free direction estimates with smaller scatter. The mean values of the direction estimator $R$ is shown in black `$+$', `$\Yup$' and `$\vartriangle$' markers for $\ell_\text{min}$ of 2, 10 and 20 respectively. These are very close to the actual direction used in the simulation denoted by the black `$\times$'.

For the analysis in the paper, we will use $\ell_\text{min}$ of 10, 20 and 40. The primary source of foreground contamination for CMB polarization measurements are polarized dust emissions. The E and B mode power spectra for the dust emissions have a power law behaviour $C^{XX}_\ell \propto \ell^{\alpha_{XX}}$ with a negative spectral index $\alpha_{XX}$ \citep{Adam:2014}. For low-$\ell$ the polarized dust foreground is large. Thus including low-$\ell$ data will also include additional uncertainty due to foreground removal which are larger at these scales. Along with the possible foreground contamination for modes with $\ell<40$, for the Planck 2018 polarization data, we know that there are systematic residuals present mainly due to calibration error. Thus a lower $\ell_\text{min}$ cut makes the estimation sensitive to these other issues over and above the usual cosmic variance.  We discussed previously the dipole modulation effect is scale dependent. Since the amplitude is larger for the largest angular scales the inclusion of the low-$\ell$ modes in the polarization analysis may be important in the search for this signal.
\begin{table}
    \centering
	\begin{tabular}{c  r r }
	    \hline
        $\ell_\text{min}$ & 10 & 40 \\
        \hline 
        \noalign{\vskip 0.1cm}
        $\beta_0$ & $0.0148 \pm 0.0114$ & $0.0140 \pm 0.0106$\\
        $\beta_1$ & $-0.00153 \pm 0.00113$ & $-0.00170 \pm 0.00108$\\
        $\beta_2$ & $0.000051 \pm 0.000024$ & $0.000062 \pm 0.000024$\\
    \hline
    \end{tabular}
    \caption{Bias correction model parameters for $\ell_\text{min}$ of 10 and 40, obtained with CORE-like noise.}
    \label{tab:bias_params}
\end{table}

\subsection{Estimator performance and bias correction}
We use simulated CMB maps which are modulated as described in \ref{Sec:SimMod}. Noise maps, prepared with CORE specifications, as discussed in \ref{Sec:SimNoise}, are added to the simulated maps. We prepared $P^2$ maps as discussed above and estimated the direction $\boldsymbol{\hat \lambda}$ with $R$ and $D$ estimators as described above. The scatter plot for the $R$ estimator is shown in the top row of figure \ref{Fig:COREforecast} for two cases of $\ell_\text{min}=10$ and $\ell_\text{min}=40$. The corresponding
plot for $D$ estimator is similar. We note that both the $R$ and $D$ estimators maximize along exactly same directions. For masked sky the mean reconstructed direction is $(231^\circ \pm 13^\circ , -13^\circ \pm 12^\circ)$ for $\ell_\text{min}=10$ and $(232^\circ \pm 17^\circ , -14^\circ \pm 14^\circ)$ for $\ell_\text{min}=40$ with the $R$ estimators. We can see that both the estimators have a negligible bias. The $R$ and $D$ estimators can be assumed to be unbiased estimators for the direction of modulation. There is $\sim 15^\circ$ error in the estimated direction coordinates. 

\begin{table*}
    \centering
    \begin{tabular}{lcccccccc}
    \hline
         \noalign{\vskip 0.1cm}
         \multirow{2}{*}{\textsc{Map}}& \multirow{2}{*}{$\ell_\text{min}$}& \multirow{2}{*}{$\ell_\text{max}$} & \multicolumn{3}{c}{R estimator} & \multicolumn{3}{c}{D estimator}\\\cline{4-9}
          \noalign{\vskip 0.1cm}
         & & & $A$ & $\boldsymbol{\hat \lambda}$ & p-value & $A$ & $\boldsymbol{\hat \lambda}$ & p-value\\
     \hline
      \noalign{\vskip 0.1cm}
         \multirow{3}{*}{Commander} & 10 & 128 & $0.572 \pm 0.099$ & $(l=352^\circ  b=-13^\circ)$& $0.007$ & $0.565 \pm 0.098$ & $(l=350^\circ  b=-14^\circ)$&$0.007$ \\
          & 20 & 128 & $0.439 \pm 0.081$ & $(l=346^\circ  b=-18^\circ)$& $< 1/300$ & $0.435 \pm 0.081$ & $(l=347^\circ  b=-20^\circ)$& $0.003$\\
          & 40 & 128 & $0.336 \pm 0.053$ & $(l=352^\circ  b=-13^\circ)$&$< 1/300$ & $0.336 \pm 0.053$ & $(l=352^\circ  b=-13^\circ)$&$< 1/300$\\
         \hline
          \noalign{\vskip 0.1cm}
         \multirow{3}{*}{SMICA} & 10 & 128 & $0.181 \pm 0.074$ & $(l=228^\circ  b=-1^\circ)$& $0.45$ & $0.182 \pm 0.074$ & $(l=226^\circ  b=-2^\circ)$&0.44\\
          & 20 & 128 & $0.099 \pm 0.066$ & $(l=232^\circ  b=-30^\circ)$& $0.72$ & $0.098 \pm 0.066$ & $(l=233^\circ  b=-31^\circ)$&$0.73$\\
          & 40 & 128 & $0.062 \pm 0.046$ & $(l=53^\circ  b=75^\circ)$&$0.77$ & $0.062 \pm 0.046$ & $(l=58^\circ  b=75^\circ)$&$0.76$\\
         \hline
         \noalign{\vskip 0.1cm}
         \multirow{3}{*}{SEVEM} & 10 & 128 & $0.627 \pm 0.102$ & $(l=352^\circ  b=-21^\circ)$& $< 1/300$ & $0.620 \pm 0.103$ & $(l=350^\circ  b=-22^\circ)$& $< 1/300$\\
          & 20 & 128 & $0.818 \pm 0.095$ & $(l=346^\circ  b=-4^\circ)$& $< 1/300$ & $0.813 \pm 0.096$ & $(l=344^\circ  b=-5^\circ)$& $< 1/300$\\
          & 40 & 128 & $0.628 \pm 0.062$ & $(l=345^\circ  b=0^\circ)$& $< 1/300$ & $0.626 \pm 0.062$ & $(l=344^\circ  b=-1^\circ)$& $< 1/300$\\
         \hline
         \noalign{\vskip 0.1cm}
         \multirow{3}{*}{NILC} & 10 & 128 & $0.464 \pm 0.169$ & $(l=332^\circ  b=-16^\circ)$& $0.65$ & $0.461 \pm 0.170$ & $(l=333^\circ  b=-17^\circ)$& $0.66$\\
          & 20 & 128 & $0.246 \pm 0.137$ & $(l=329^\circ  b=-29^\circ)$& $0.97$ & $0.248\pm 0.137$ & $(l=332^\circ  b=-29^\circ)$& $0.97$\\
          & 40 & 128 & $0.319 \pm 0.084$ & $(l=2^\circ  b=-1^\circ)$& $0.66$ & $0.318 \pm 0.084$ & $(l=3^\circ  b=-2^\circ)$& $0.66$\\
     \hline
    \end{tabular}
    \caption{Planck 2018 polarization results for Commander, SMICA, SEVEM and NILC maps at \texttt{NSIDE} = 
    64. The bias uncorrected amplitude and the direction of modulation is provided for analysis both $R$ and $D$ estimators. We also provide the p-value for the corresponding result obtained from FFP10 simulations. The errors in the estimated amplitude and direction are discussed in the text.}
    \label{tab:Planck64results}
\end{table*}

The amplitude estimator defined in Eq. \eqref{Eq:AestPS} is a biased estimator. The bias varies with the modulation amplitude. To study and characterize the bias in $\hat A$ we performed simulations with different modulation amplitudes from $0.0$ to $1.0$ in steps of $0.02$. For each input modulation amplitude, we performed 150 simulations using the process described in section \ref{Sec:Sim} for CORE-like noise. We estimated the amplitude using the pipeline described above and from these estimates calculated the mean bias corresponding to the mean value of $\hat A$. The error in the bias is calculated from the standard deviation of $\hat A$. The plot of the bias values versus the $\hat A$ mean is shown in figure \ref{Fig:bias_var}. The bias is large for no modulation or small modulation amplitudes. It decreases and becomes almost negligible for larger modulation amplitudes. The bias decreases with a decrease in the skewness of the underlying distribution. 

We use the simulated data to derive a parametric model for our $\hat A$ estimator bias. Our assumed bias model $b(\hat A)$ is: 
\begin{equation}
    b(\hat A) = \beta_0 + \beta_1 \hat A^{-1} + \beta_2 \hat A^{-2}.
    \label{Eq:bias_model}
\end{equation}
We use our simulated data to fit the parameters $\beta_0$, $\beta_1$ and $\beta_2$. The best-fit parameter values for the two cases of $\ell_\text{min}=10$ and $\ell_\text{min}=40$ are given in table \ref{tab:bias_params}. We use this parametric bias model in our forecast for a CORE-like experiment.  We calculate the bias-corrected amplitude as $\hat A_c = \hat A - b(\hat A)$. During implementation on the null distribution, we note that this bias correction model should be applied only to the mean of the distribution to estimate the bias for all samples of the null simulation. This is because the bias correction model is a best-fit obtained with the mean of the $\hat A$ distribution for different input amplitude. The model is not valid below the mean of null/isotropic distribution. The model will overcompensate for small values of $\hat A$ which are below the mean amplitude of the null distribution. The skewness of the null distribution also affects the bias correction when applied to individual samples of the null distribution. When the input amplitude increases, the bias becomes negligible. Thus for larger values of modulation amplitude, the bias correction can all together be avoided. We would point out that the bias model is dependent on our noise model. It is also understandable that the bias model that is assumed here would work fairly well for $10 \le \ell_\text{min} \le 40$. We can also see from figure \ref{Fig:bias_var} that the bias is slightly higher at small amplitudes and becomes negligible faster for $\ell_\text{min}=40$ than for the $\ell_\text{min}=10$. The $\ell_\text{min}=10$ case has slightly more residual bias at higher amplitudes.

\begin{figure*}
    \centering
    \subfloat{\includegraphics[width=0.315\textwidth]{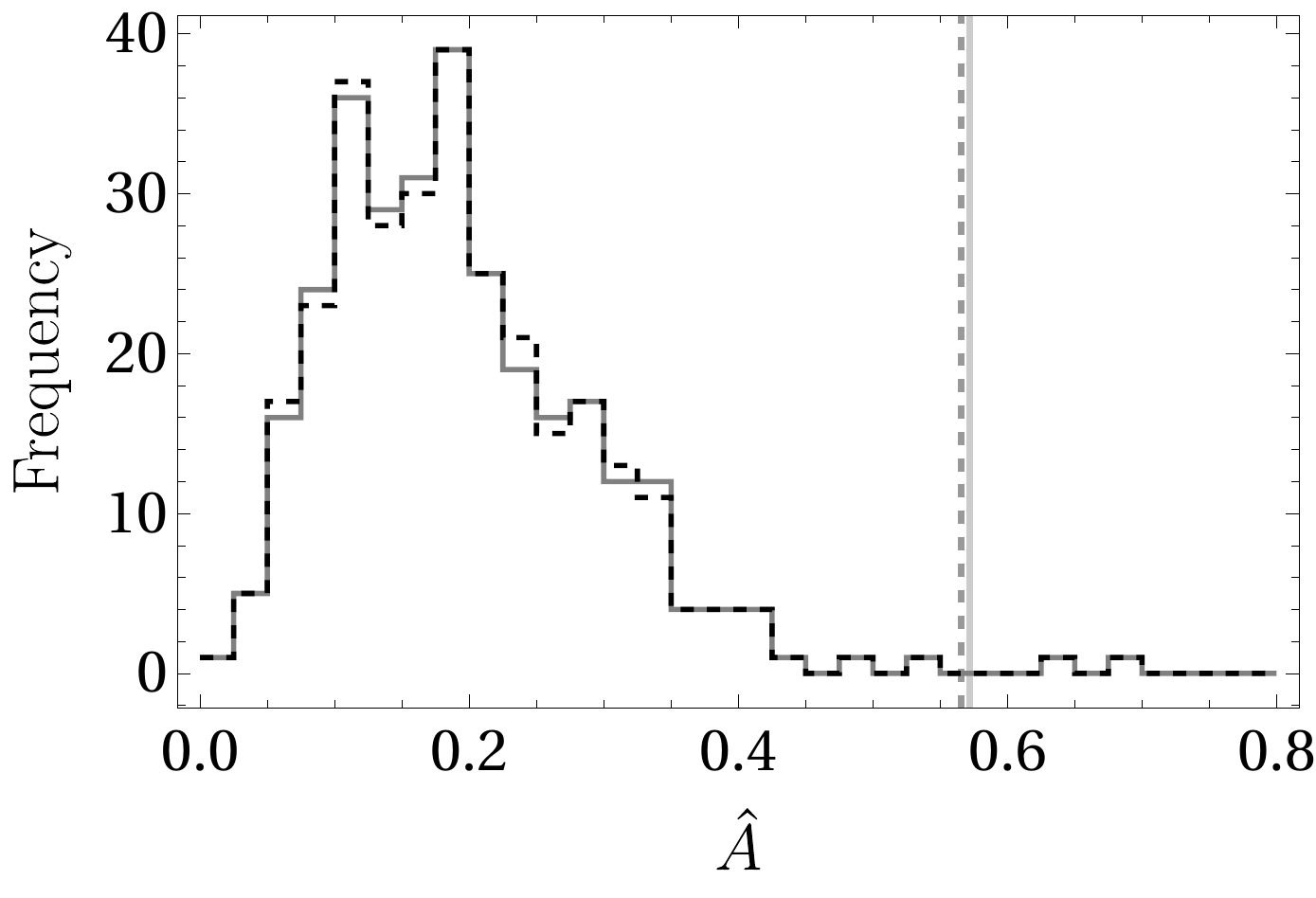}}\quad
    \subfloat{\includegraphics[width=0.315\textwidth]{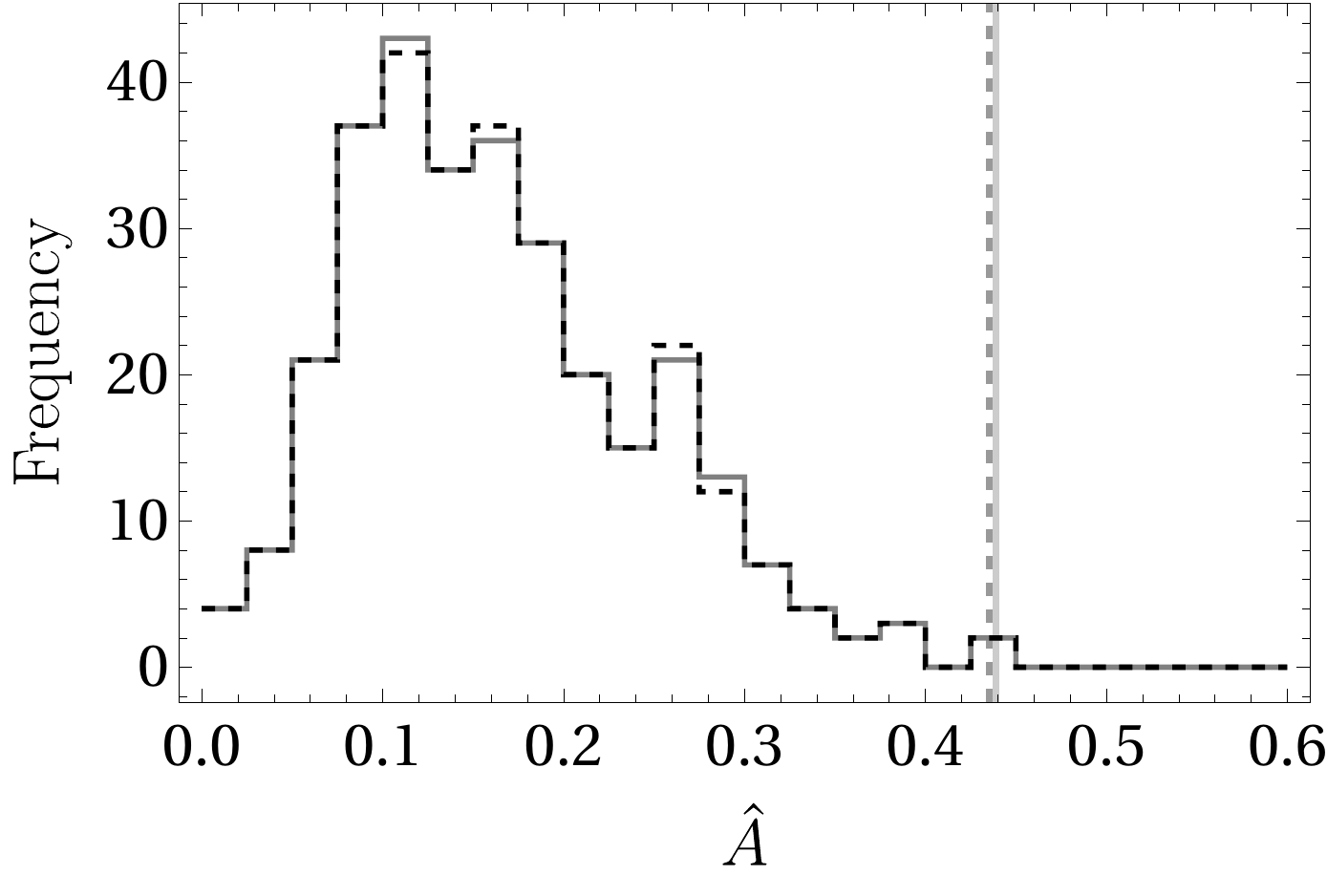}}\quad
    \subfloat{\includegraphics[width=0.315\textwidth]{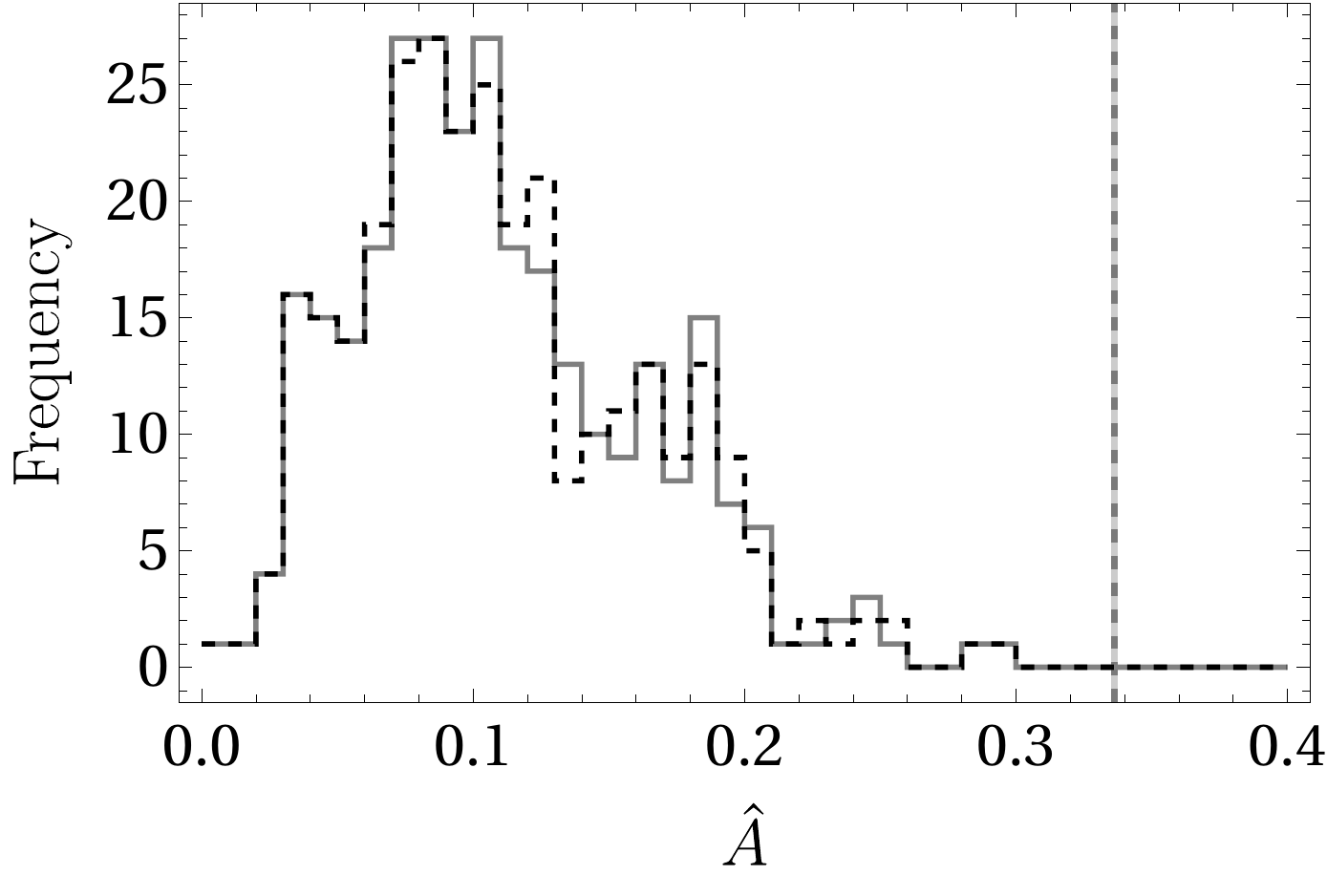}}\\
    \subfloat{\includegraphics[width=0.315\textwidth]{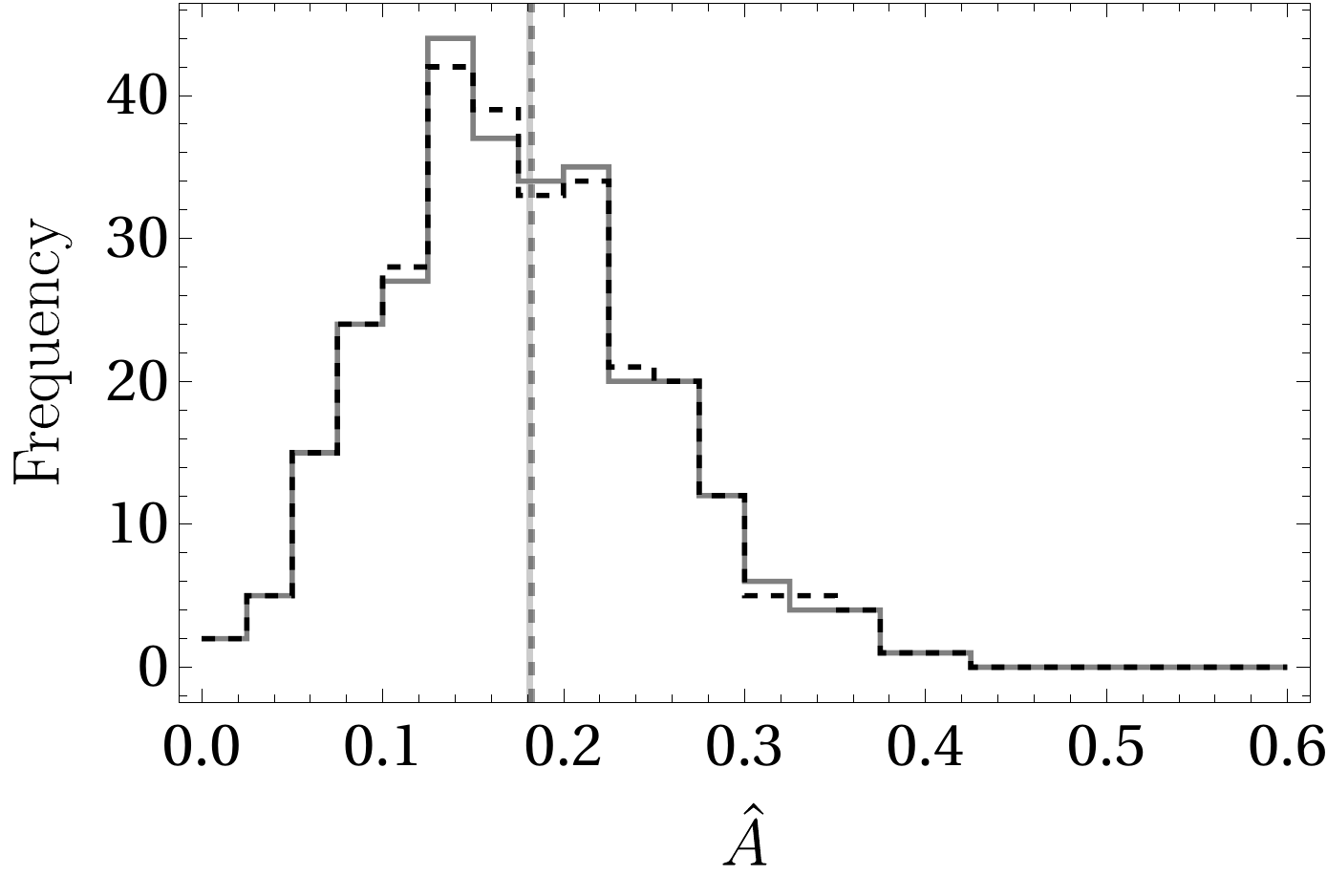}}\quad
    \subfloat{\includegraphics[width=0.315\textwidth]{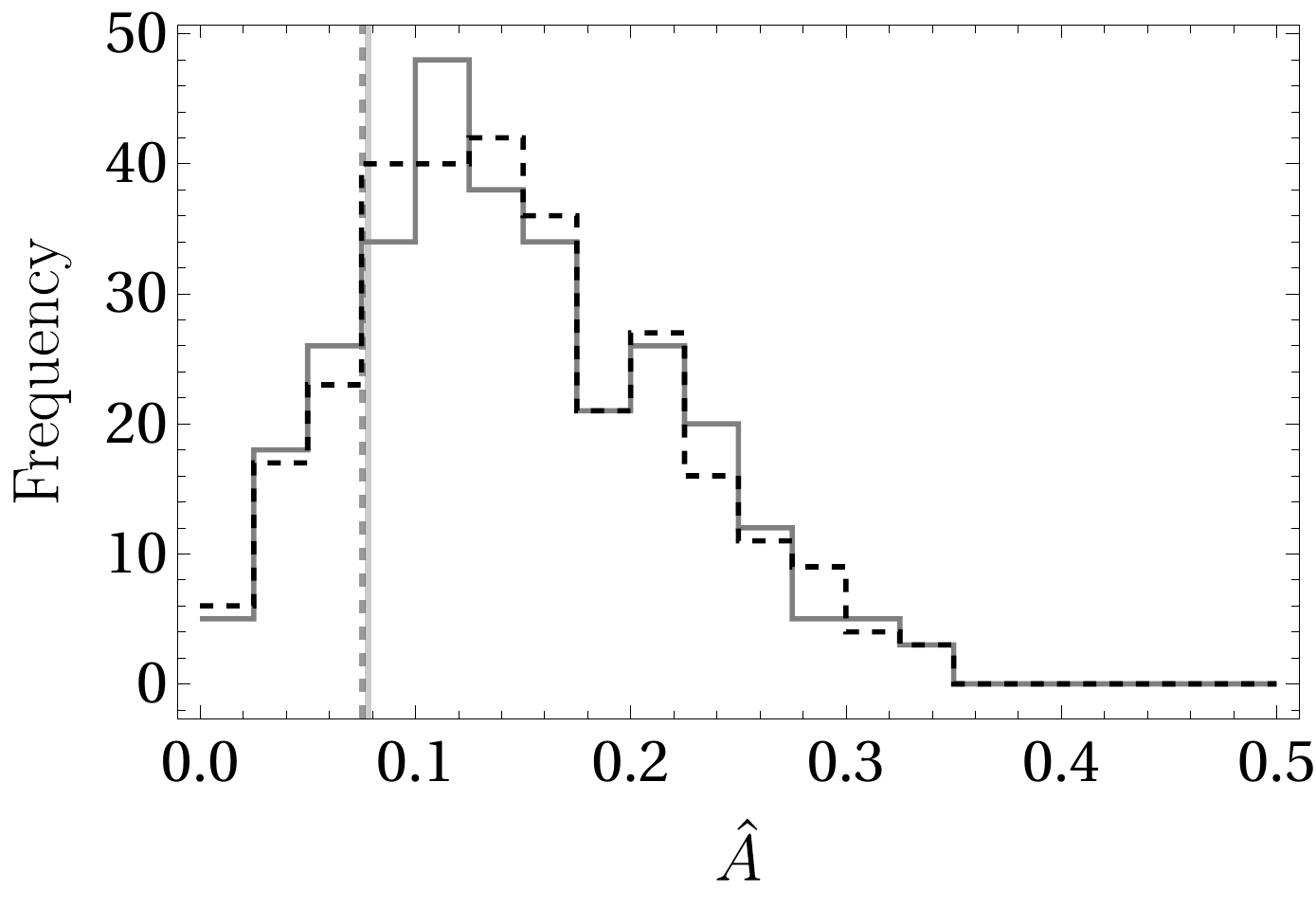}}\quad
    \subfloat{\includegraphics[width=0.315\textwidth]{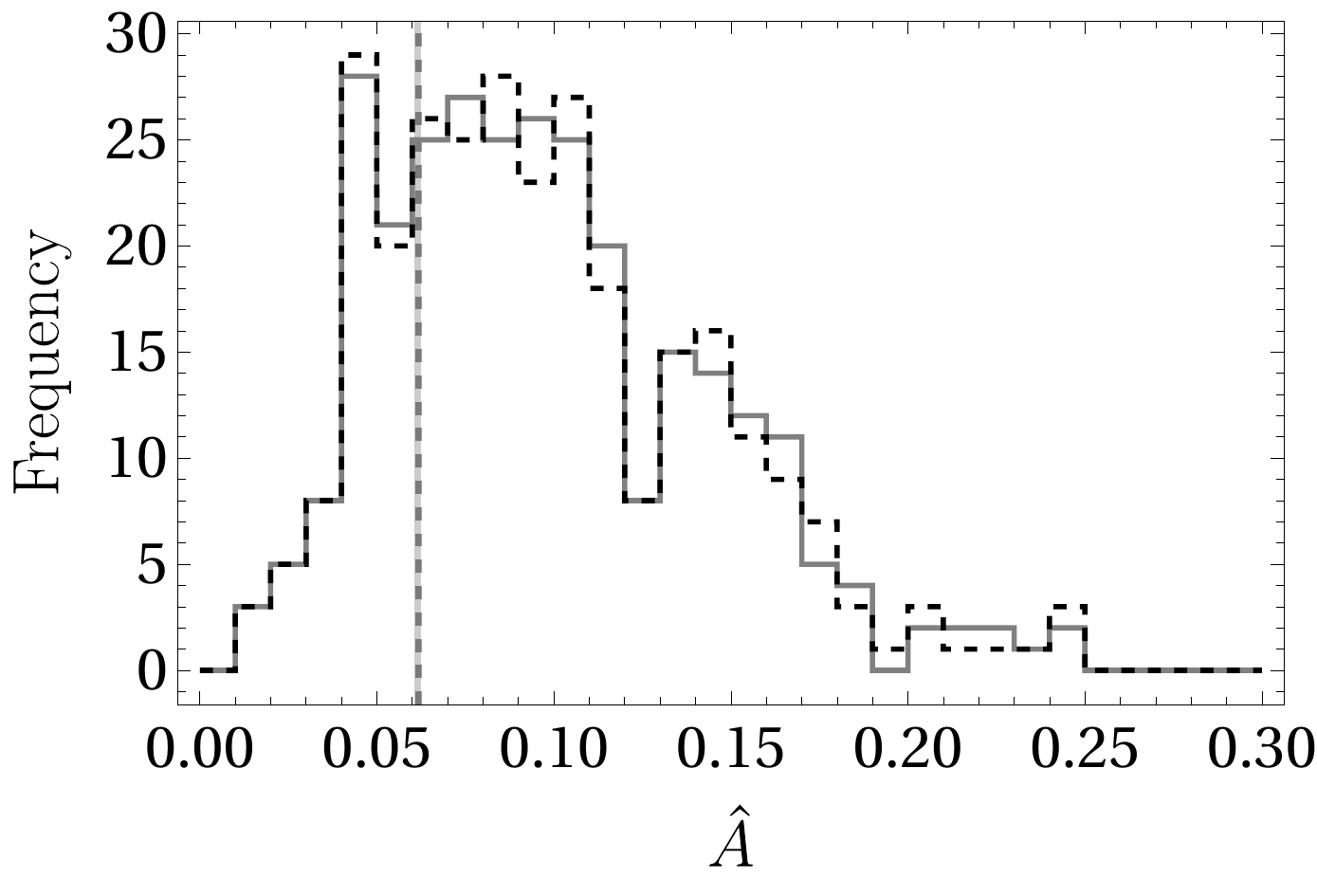}}\\
    \subfloat{\includegraphics[width=0.315\textwidth]{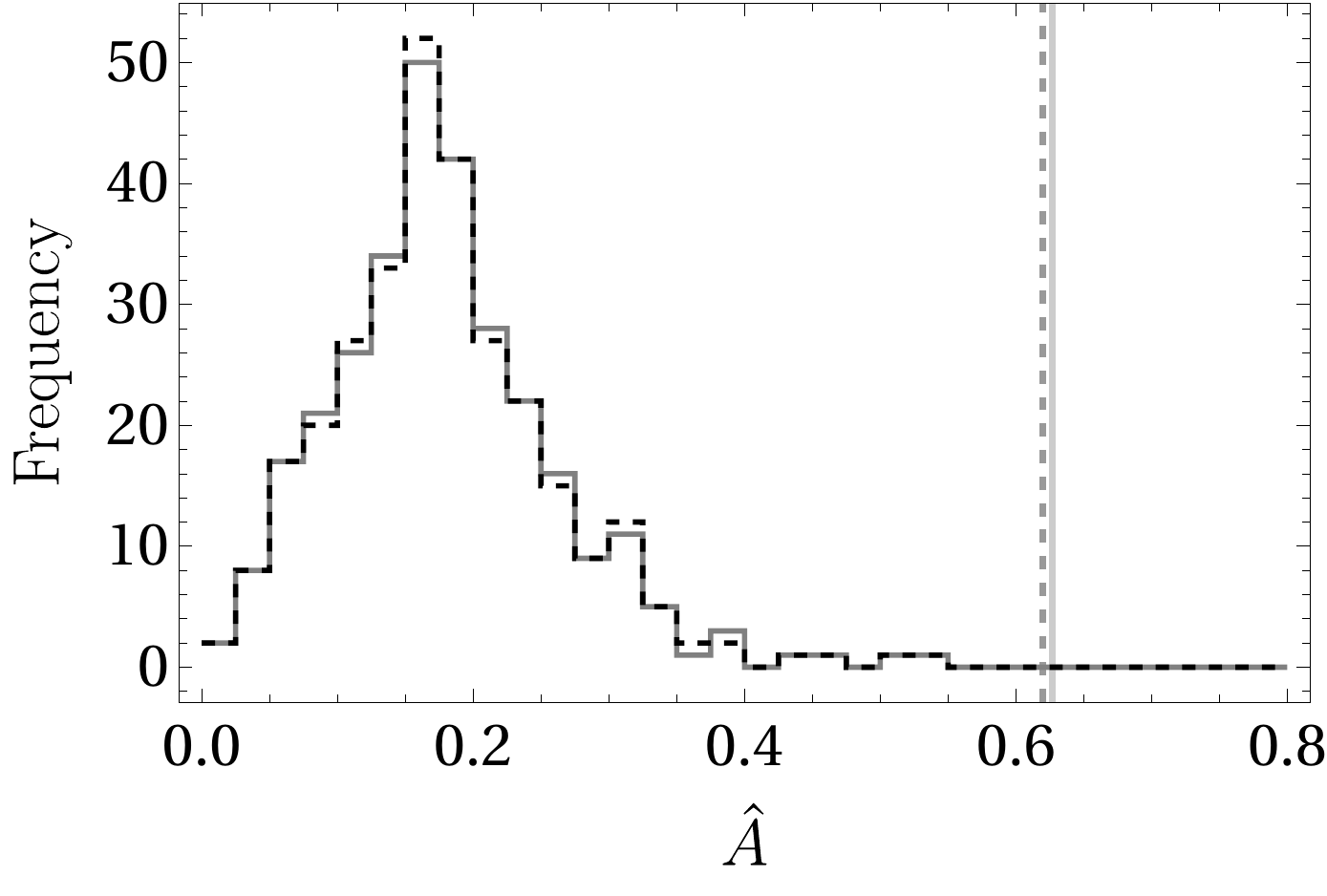}}\quad
    \subfloat{\includegraphics[width=0.315\textwidth]{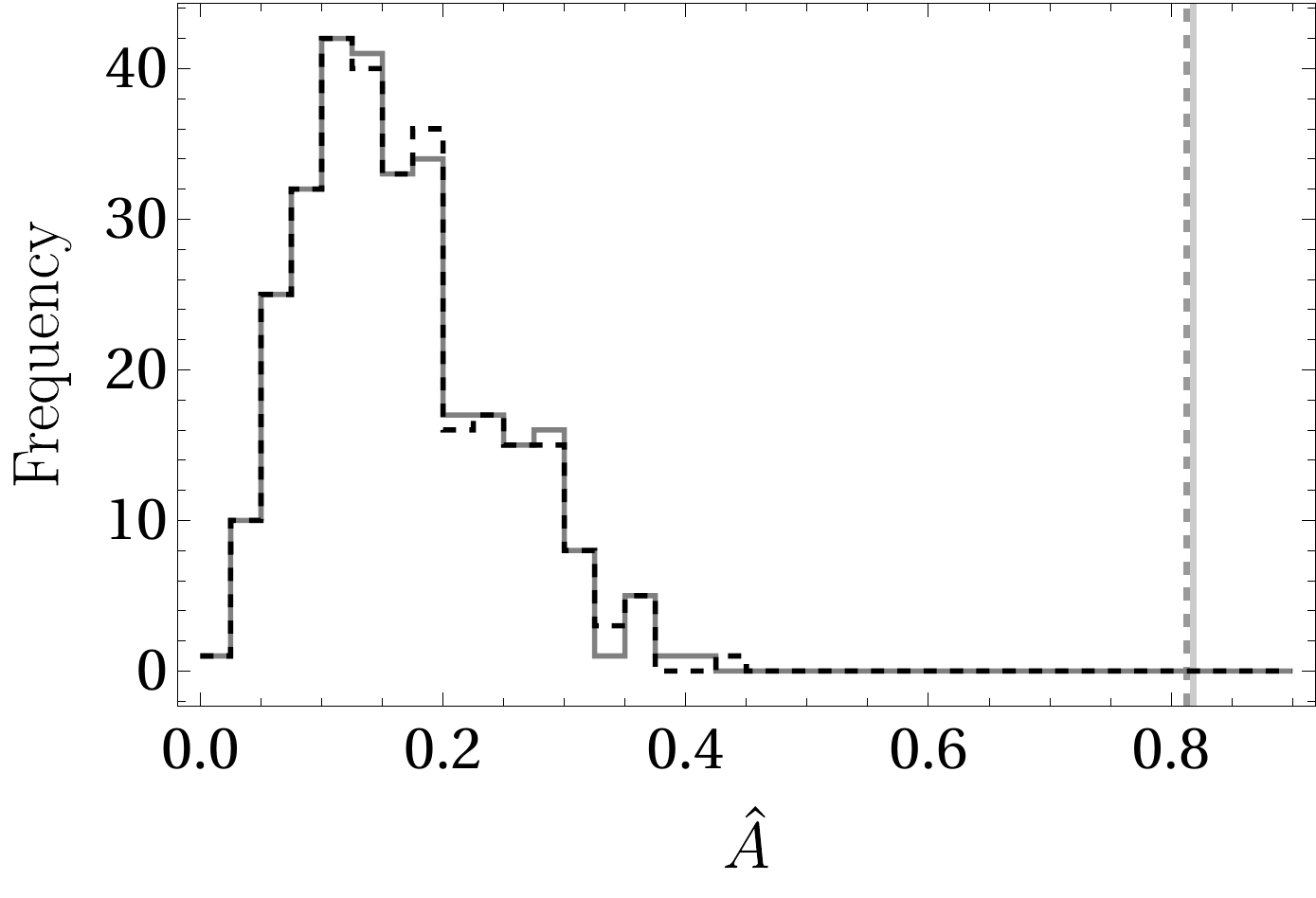}}\quad
    \subfloat{\includegraphics[width=0.315\textwidth]{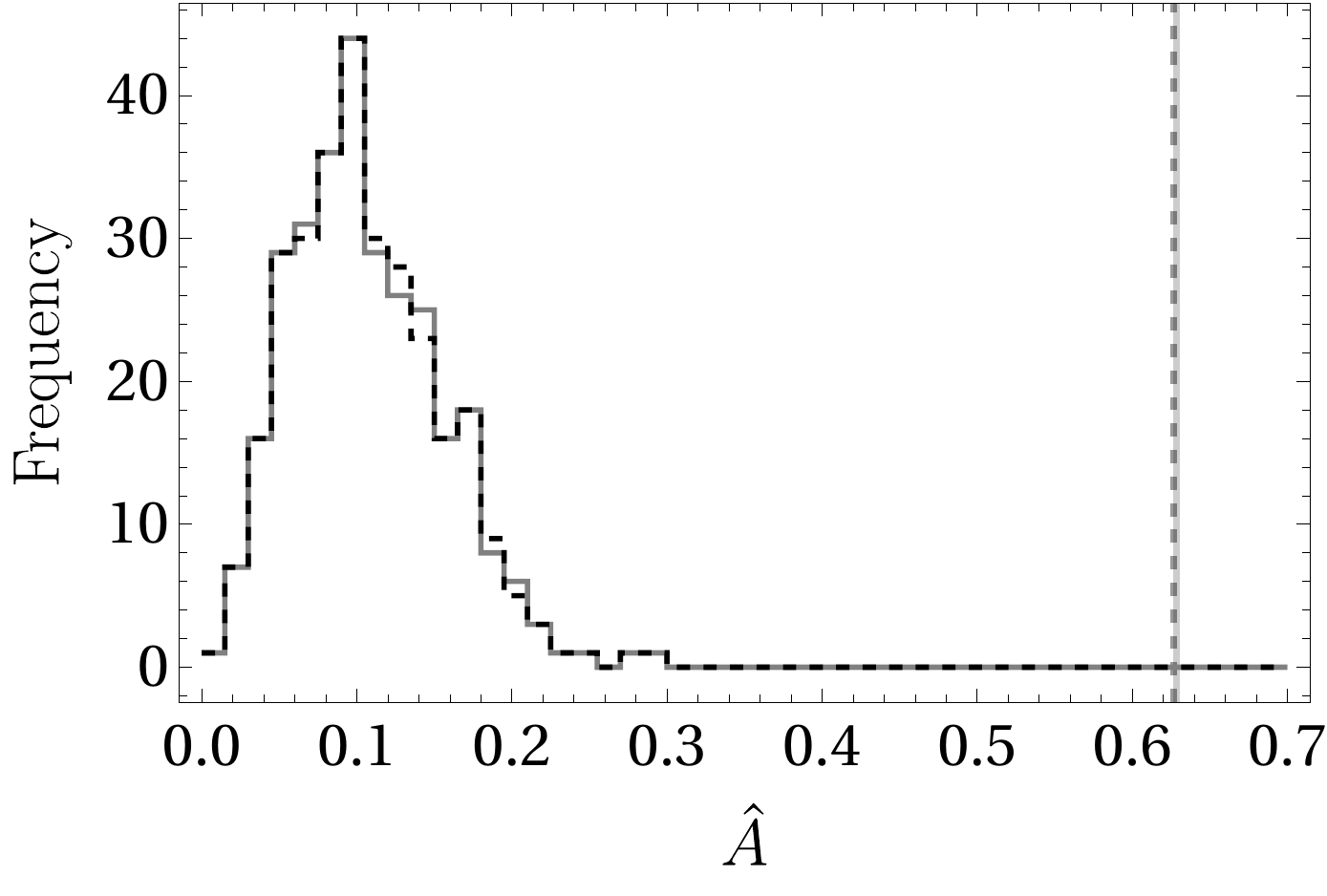}}\\
    \subfloat{\includegraphics[width=0.315\textwidth]{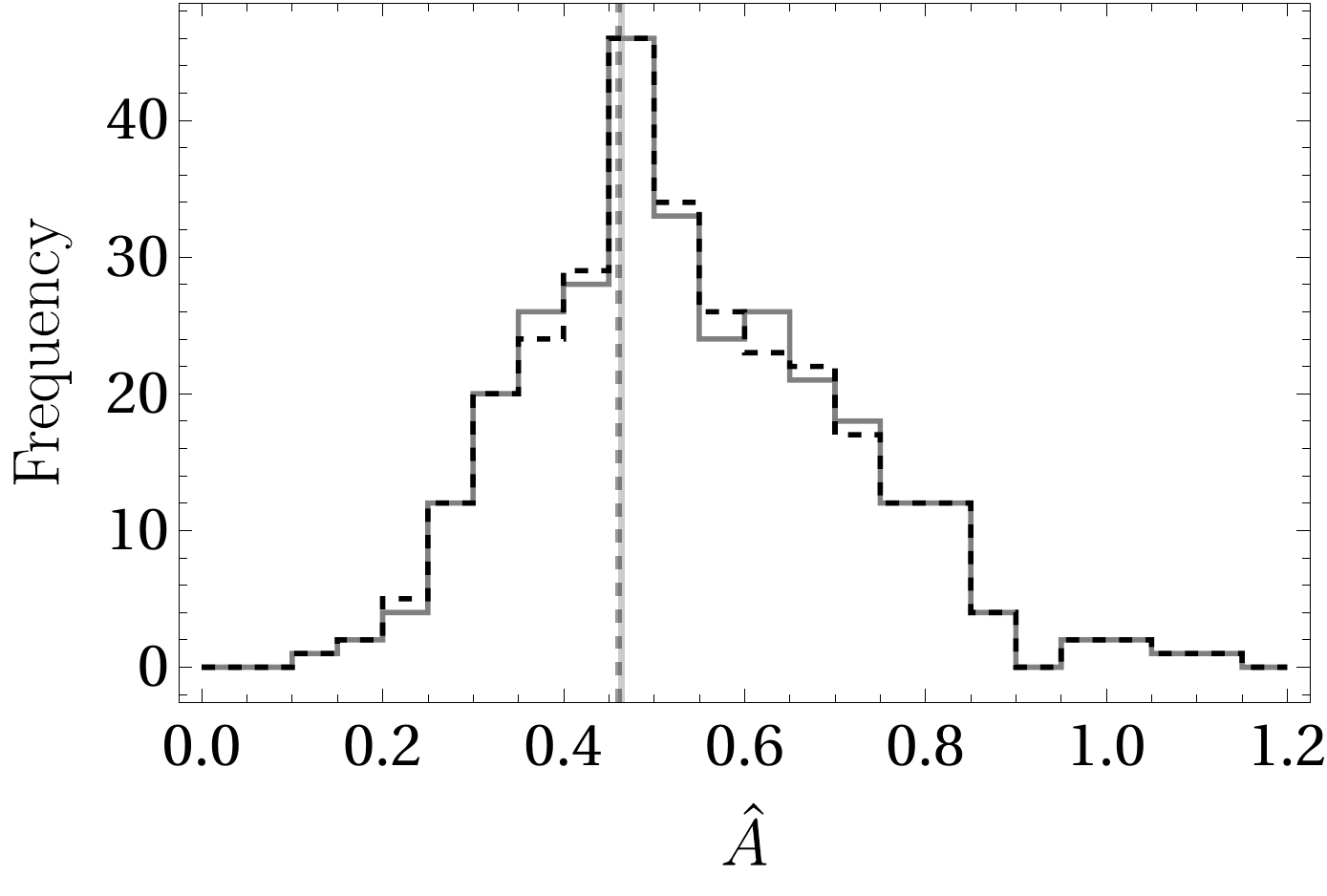}}\quad
    \subfloat{\includegraphics[width=0.315\textwidth]{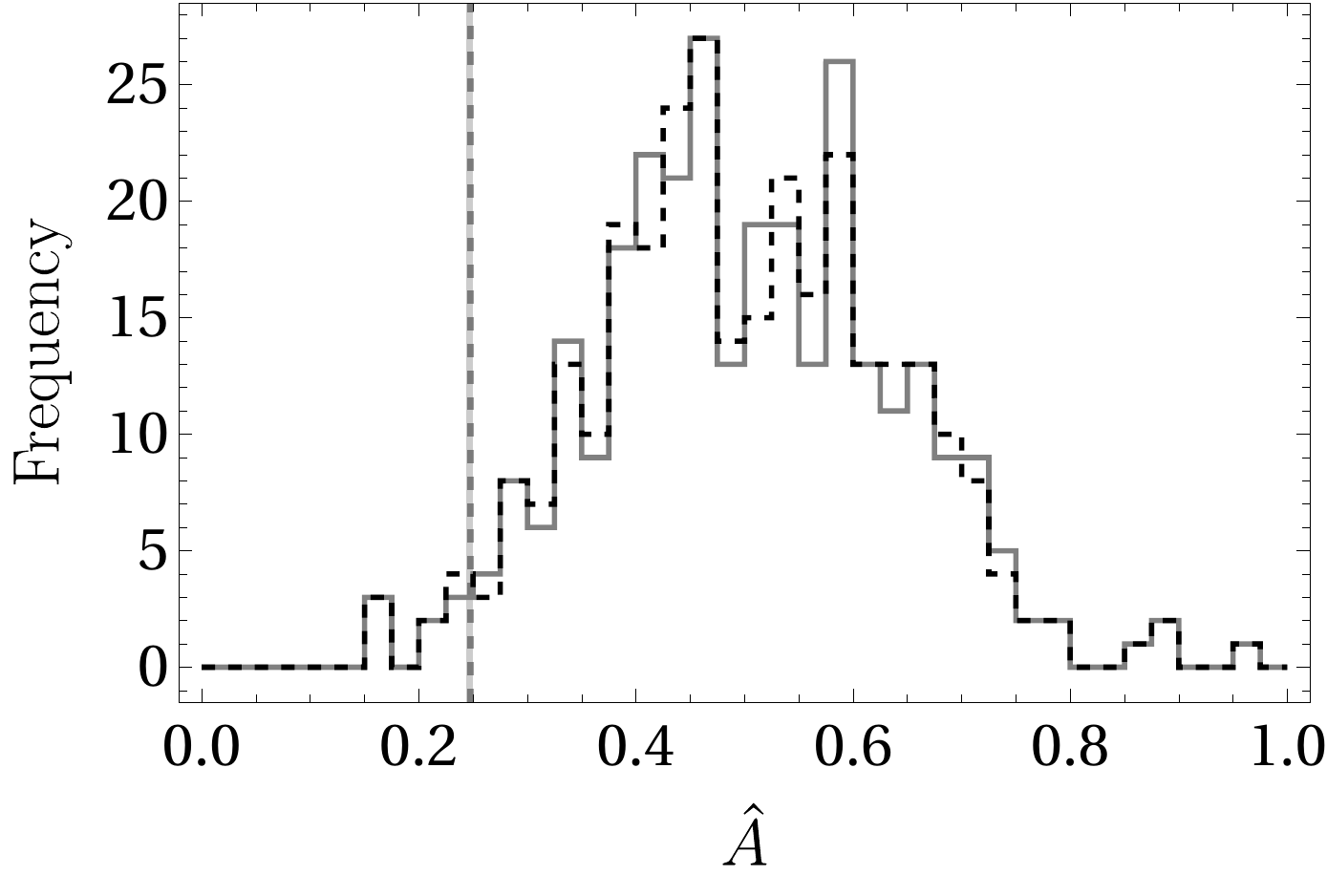}}\quad
    \subfloat{\includegraphics[width=0.315\textwidth]{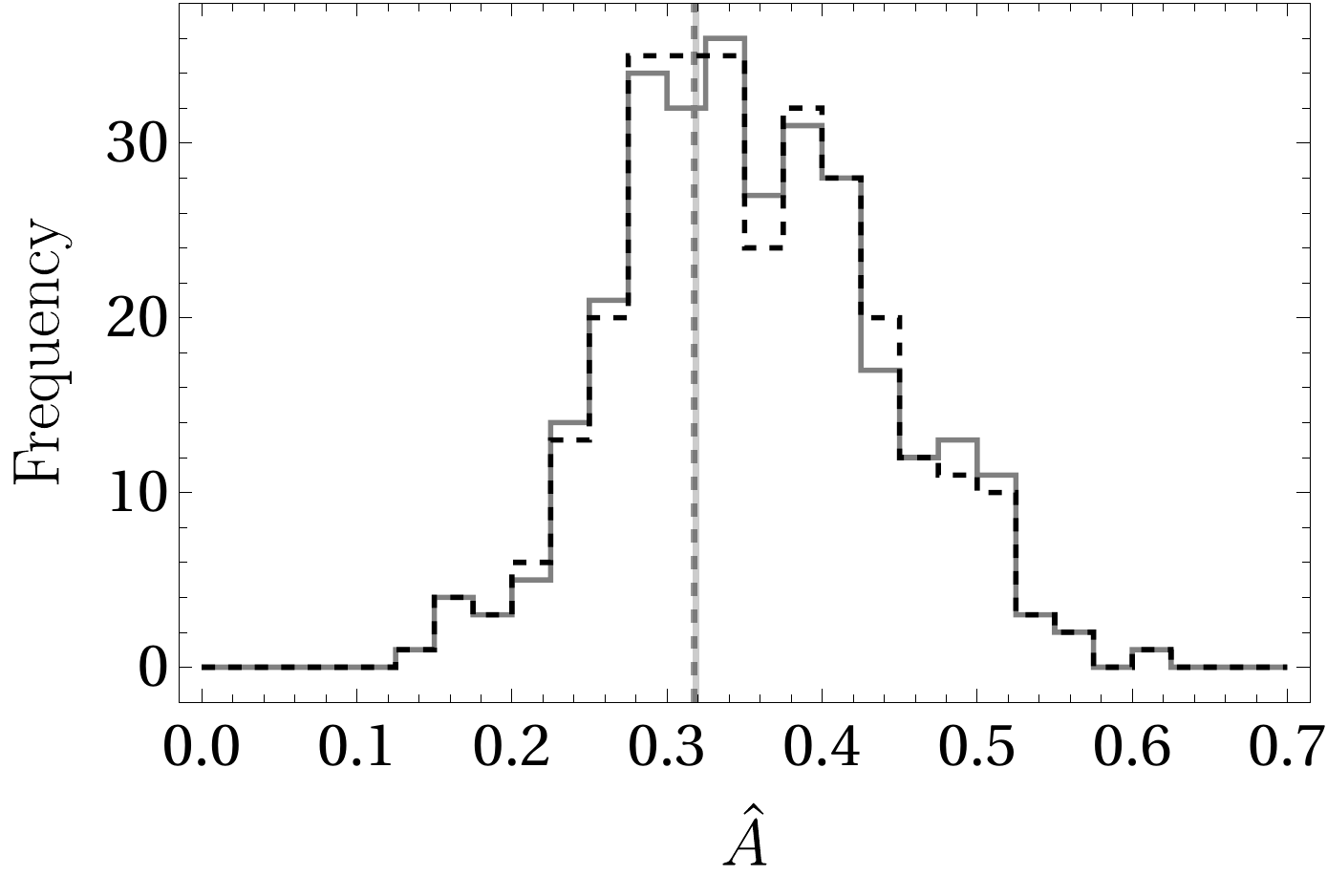}}\\
    \caption{We plot the histogram for the recovered amplitude with the $\hat A$ estimator with the FFP10 simulations for different component separation methods. We analysed 300 CMB and noise maps for each component separation method. The rows one, two, three and four are for component separation methods Commander, SMICA, SEVEM and NILC respectively. The columns one, two and three are for $10\le \ell \le 64$, $20\le \ell \le 64$ and $40\le \ell \le 128$ ranges respectively. The solid grey histogram is for amplitude estimate along the direction estimated by $R$ estimator and the dashed, black histogram is for amplitude estimated along the direction given by maximising the $D$ estimator. The solid, grey line and the dashed, black lines are the amplitude estimates of the 2018 Planck $|P|^2$ maps along the direction estimated by the $R$ and $D$ estimators respectively.}
    \label{Fig:Planck2018hist}
\end{figure*}

\subsection{Forecast for CORE-like experiment}
In figure \ref{Fig:COREforecast} we present our forecast for a CMB polarization modulation with an amplitude of $0.07$ with CORE-like instrumental noise. We simulated these results with $\ell_\text{min}=10$ (shown on the left of the figure) and with $\ell_\text{min}=40$ (shown on the right of the figure). The direction observations with a masked sky is shown in the top row. The results shown are obtained by maximising the $R$ statistic. The amplitude estimator forecast is shown in the bottom row. The dashed, black histogram is for the bias-corrected null distribution, where we have corrected the distribution for the bias corresponding to the mean amplitude of the null distribution. We show the bias-uncorrected null distribution in the dotted, grey histogram. The histogram shown with black, solid line is for the modulated case without bias correction. With bias correction, there is very small change in this histogram. The solid black line indicates the mean of the modulated distribution with the grey band showing 1$\sigma$ bound. It is clear that the amplitude estimator would work well to detect the amplitude with or without bias correction when the amplitude is $0.07$. The significance of the detection increases with bias correction. 
From the plots in figure \ref{Fig:COREforecast} it is apparent that a $7 \%$ dipole modulation of the kind observed in CMB temperature fluctuations should be detectable at over $3\sigma$ with a CORE-like full sky experiment, using our estimators with proper bias correction. From the figure \ref{Fig:COREforecast} it is clear that the modulation signal is clearly distinct from the isotropic distribution even in absence of any bias corrections. The mean of the modulated distribution is 4.8$\sigma$ deviations away from the mean of the bias uncorrected null distribution. So the mean is clearly detectable at $> 3\sigma$. While a $\mp 1\sigma$ variation about the mean of the modulation distribution shows a deviation of $2.7-5.8\sigma$ from the mean of the bias uncorrected null distribution. Thus, even in absence of bias corrections, we should still be able to detect as modulation signal but at a lower significance. We should also be able to estimate the direction with acceptable errors. The above discussion provides justification for the use of our pixel space estimators for detecting dipole modulation in CMB polarization in future experiments.

\begin{figure*}
    \centering
    \subfloat{\includegraphics[width=0.315\textwidth]{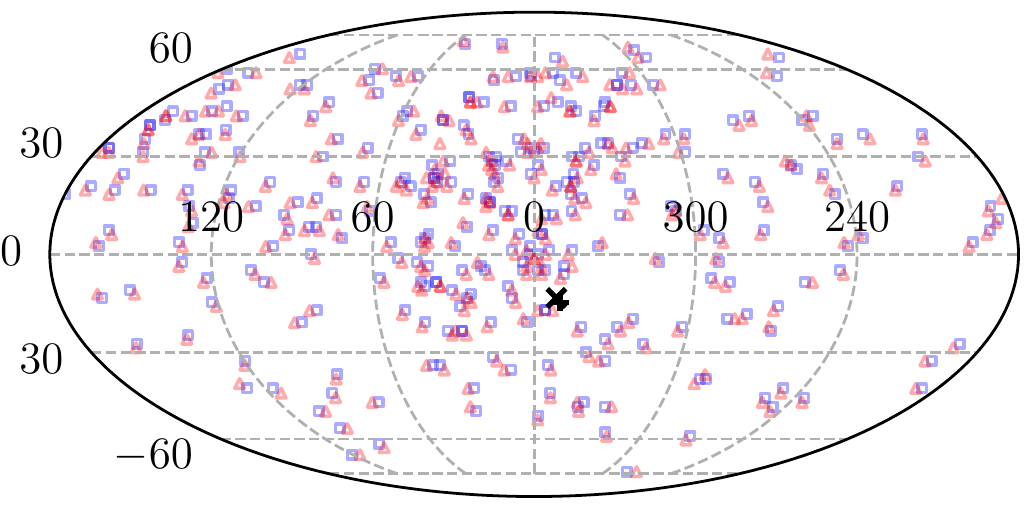}}\quad
    \subfloat{\includegraphics[width=0.315\textwidth]{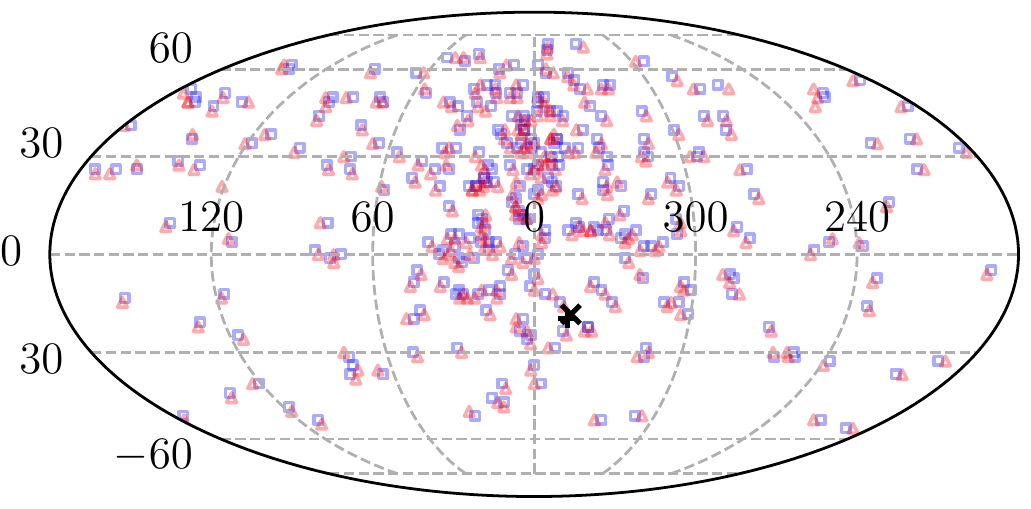}}\quad
    \subfloat{\includegraphics[width=0.315\textwidth]{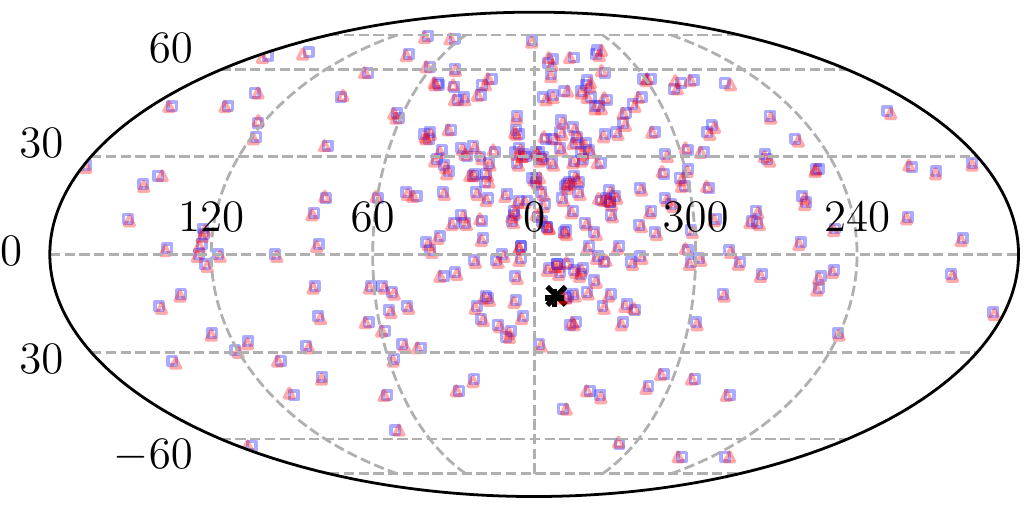}}\\
    \subfloat{\includegraphics[width=0.315\textwidth]{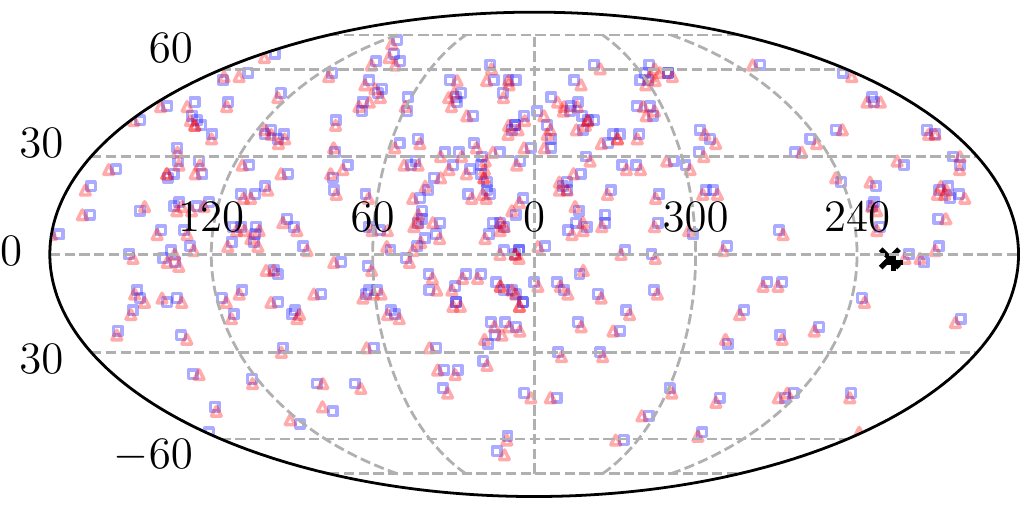}}\quad
    \subfloat{\includegraphics[width=0.315\textwidth]{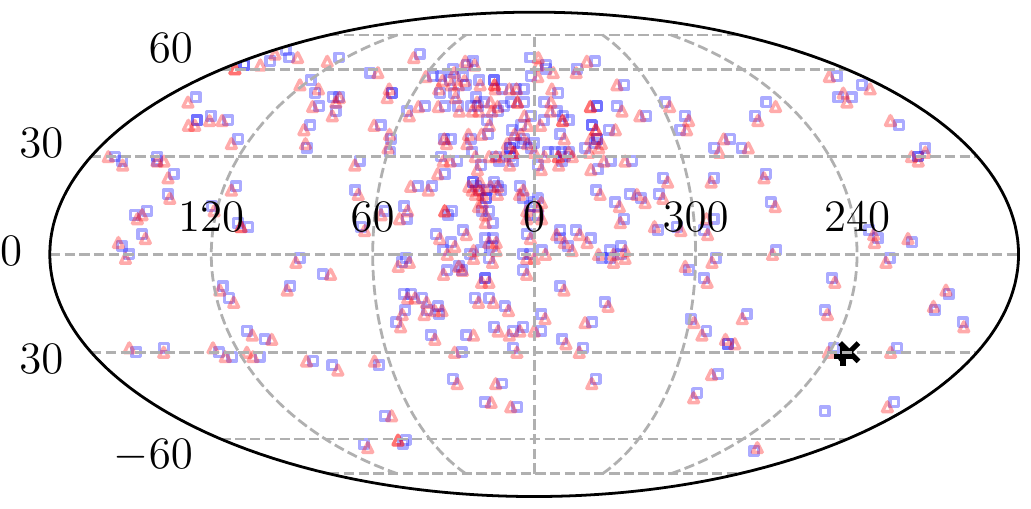}}\quad
    \subfloat{\includegraphics[width=0.315\textwidth]{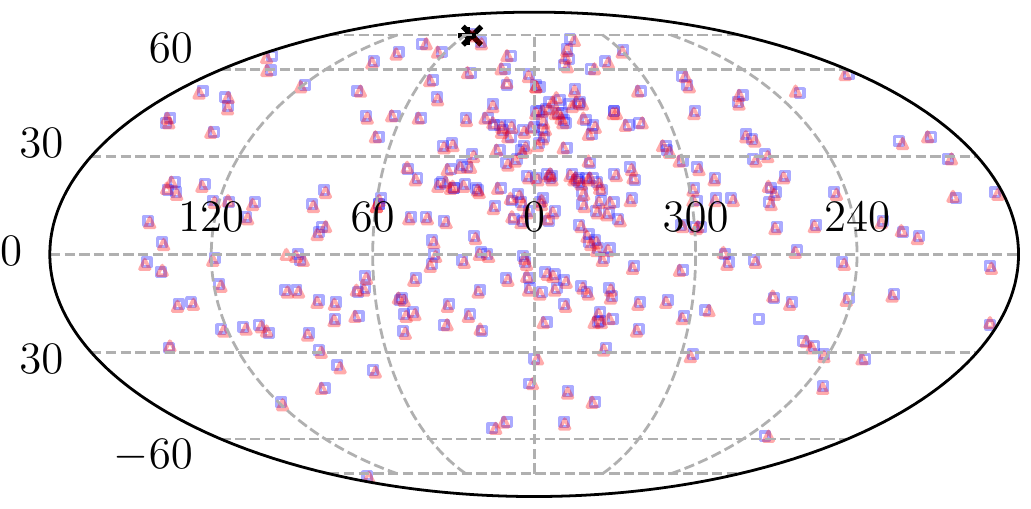}}\\
    \subfloat{\includegraphics[width=0.315\textwidth]{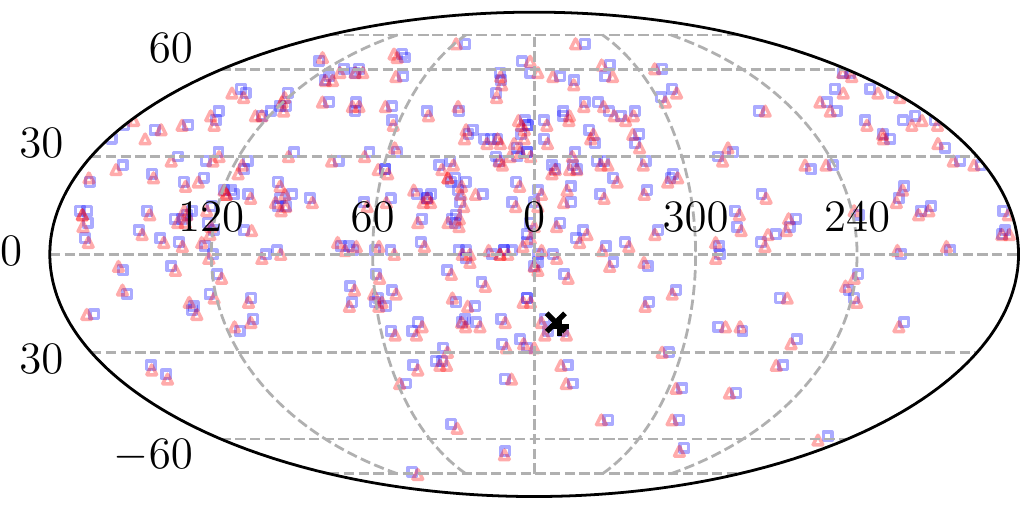}}\quad
    \subfloat{\includegraphics[width=0.315\textwidth]{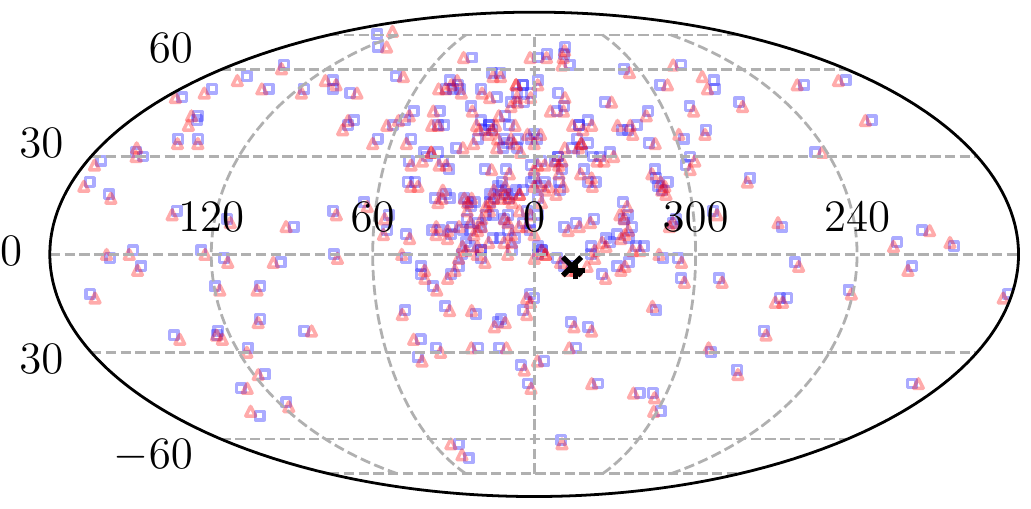}}\quad
    \subfloat{\includegraphics[width=0.315\textwidth]{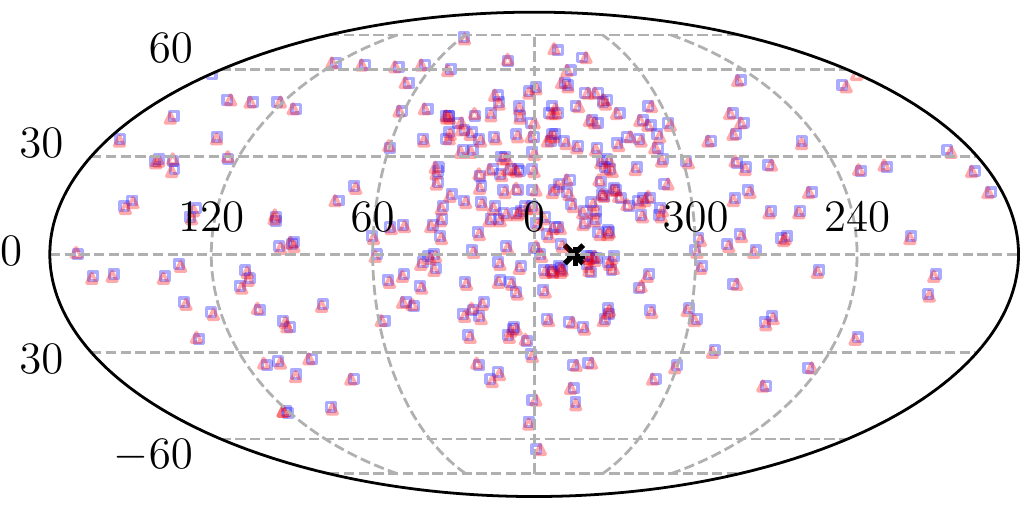}}\\
    \subfloat{\includegraphics[width=0.315\textwidth]{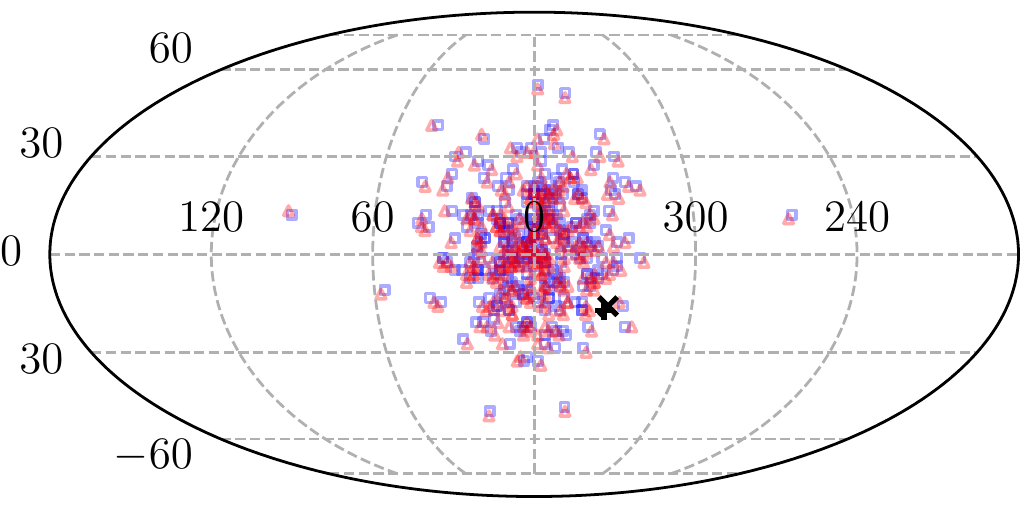}}\quad
    \subfloat{\includegraphics[width=0.315\textwidth]{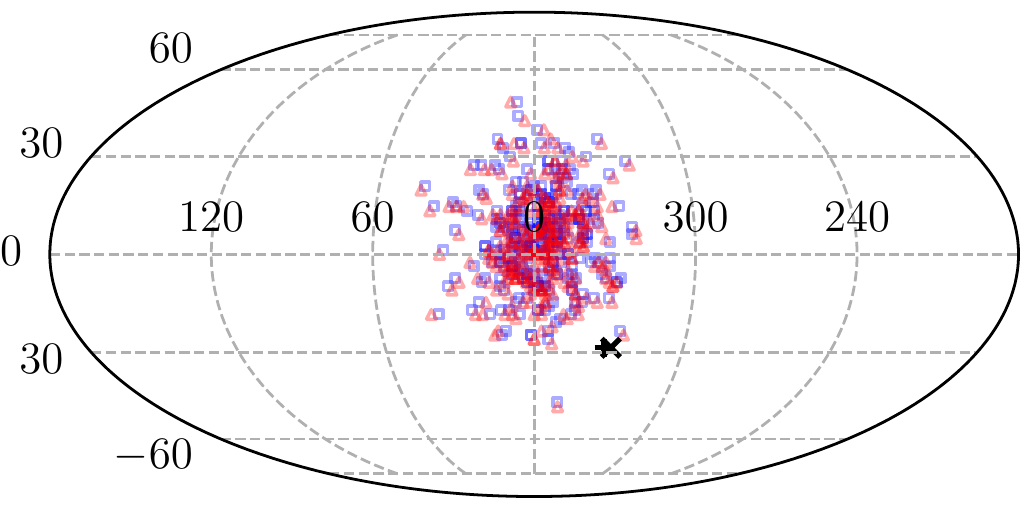}}\quad
    \subfloat{\includegraphics[width=0.315\textwidth]{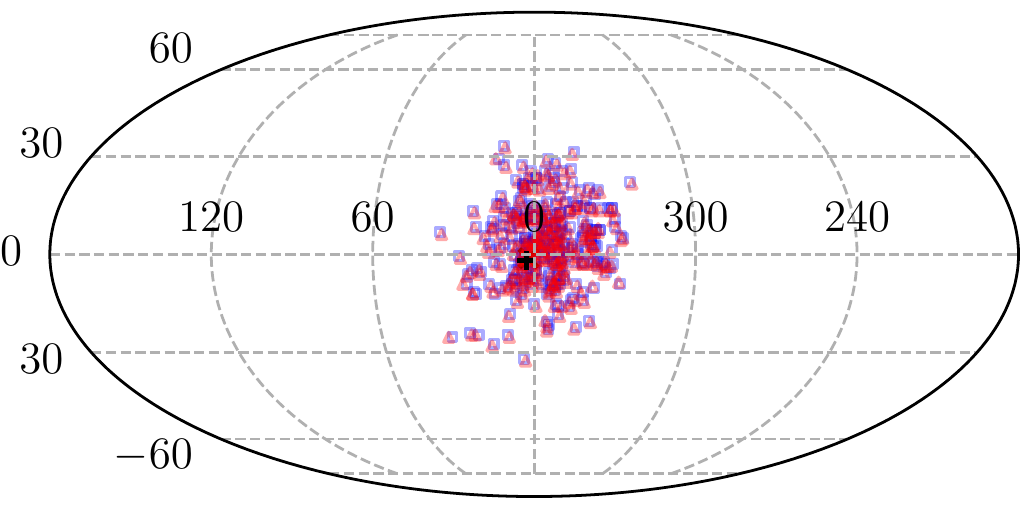}}\\
    \caption{We plot the scatter of direction recovered from the FFP10 simulations for different component separation methods. For each component separation method 300 CMB and noise maps were analysed. The rows one, two, three and four are for component separation methods Commander, SMICA, SEVEM and NILC respectively. The columns one, two and three are for $10\le \ell \le 64$, $20\le \ell \le 64$ and $40\le \ell \le 128$ ranges respectively. The blue `$\square$' represent direction estimates by maximising the $R$ statistic and the red `$\vartriangle$' represent the direction obtained by maximising the $D$ statistic. The black `$\times$' and the `$+$' are the direction estimates of the 2018 Planck $|P|^2$ maps by using the $R$ and $D$ estimators respectively, and they coincide in all plots.}
    \label{Fig:Planck2018dir}
\end{figure*}

\begin{table}
    \centering
	\begin{tabular}{c c c c }
	    \hline
        \texttt{NSIDE} & $\ell_\text{max}$ & Amplitude& $\boldsymbol{\hat \lambda}$\\
        \hline 
        \noalign{\vskip 0.1cm}
        64 & 128& 0.181 & ($228^\circ$,  $-1^\circ$)\\
        128 & 256& 0.062 & ($254^\circ$,  $27^\circ$)\\
        256 & 512& 0.042 & ($277^\circ$,  $60^\circ$)\\
        512 & 1024& 0.022 & ($20^\circ$, $66^\circ$)\\
        \hline
    \end{tabular}
    \caption{Results for bias uncorrected amplitude of dipole modulation and direction of modulation for different resolutions of the SMICA polarization map with $\ell_\text{min}=10$ using $R$ statistics to estimate the direction.}
    \label{tab:nside_vary}
\end{table}


\section{Results} \label{Sec:Results}
We have analyzed 2018 Planck CMB polarization maps cleaned by Commander, SMICA, SEVEM and NILC component separation methods. The results for the bias-uncorrected amplitude, the direction of modulation and the corresponding p-value is provided in table \ref{tab:Planck64results}. We use the FFP10 null simulations to determine the p-value for our $\hat A$ estimation. The histograms for the null distributions are shown in figure \ref{Fig:Planck2018hist}. The solid, grey and dashed, black histograms represent the $R$ statistic and $D$ statistic results respectively. The solid, grey and black, dashed lines indicate the observed amplitude from the corresponding Planck map with $R$ or $D$ estimator. The statistical errors for the amplitude estimates is calculated from the FFP10 simulations. These are also shown in table \ref{tab:Planck64results}.  There exists a strong bias or systematic effect which appears to mask any real effect which may be present in data. We do not perform any bias correction for these results. We estimated the significance of our results using 300 isotropic FFP10 simulations as described in section \ref{Sec:Sim}. The p-values quoted in table \ref{tab:Planck64results} are large for SMICA and NILC but are $< 1/300$ for SEVEM and for $\ell_\text{min}$ of 10 and 40 for Commander.  This means that most of the randomly generated samples lead to statistics values larger than those seen in data for SMICA and NILC maps but none/few would show the larger values than those measured in the SEVEM or Commander maps. We find it strange that the SEVEM and Commander maps seem to indicate a significant modulation effect even when incorporating the FFP10 noise-and-systematics in the analysis.


The scatter plots for the directions from the simulations are shown in figure \ref{Fig:Planck2018dir}. The blue `$\square$' indicate directions obtained by using the $R$ statistic and the red `$\vartriangle$' for the $D$ statistic directions. We indicate the observed direction in the data by the black `$\times$' and `$+$' for the $R$ and $D$ estimators. The first, second and third columns are for $\ell_\text{min}$ of 10, 20 and 40. The rows one to four are for Commander, SMICA, SEVEM and NILC respectively. We note strong clustering in only the NILC simulations. Several plots show clustering close to the origin. The directions recovered from the data in many cases lie very close to the galactic plane and are consistent with the FFP10 null simulations.

In table \ref{tab:nside_vary} we show the effect of varying the resolution of the $P^2$ map on the analysis. These results are for $\ell_\text{min} = 10$. We see that the measured amplitude decreases with increase in resolution. We also see that the direction of the effect migrates away from the galactic plane with an increase in resolution.

Our results from the 2018 Planck polarization data cannot be directly compared to the previous results for 2015 Planck polarization data \citep{Ghosh:2016,Aluri:2017} as the 2018 maps are cleaned by Planck by updated component separation pipelines leading to differences relevant for isotropy analysis. The directions obtained in this work is very different from those reported previosly \citep{Ghosh:2016,Aluri:2017}. For the first time we also analyse the scales $\ell$ < 40 for dipole modulation. While \citet{Ghosh:2016} does not give a result for dipole modulation amplitude, the modulation amplitude result from \citet{Aluri:2017} is much smaller than the results we obtain. This is likely due to the difference in the way the two methods deal with correlated noise when estimating the model parameter values. We do not find significant results for SMICA and NILC cleaned maps, however the SEVEM and Commander maps show a high significance modulation effect. But the direction being close to the galactic plane for both Commander and SEVEM along with the known issues of systematic residuals make it difficult to argue that the observed effect is a physical signal. Overall our estimators are sensitive to correlated noise of the kind present in Planck 2018 data due to systematic residuals. The uncorrelated noise assumption clearly is not very appropriate for Planck polarization maps. So we will err on the side of caution and suggest further investigation into the Commander and SEVEM to determine the reason for the discrepancy.

The FFP10 simulations imply that there is a very large bias in data. As an exploratory study, we attempt to correct for this bias by subtracting the mean value of the dipole modulation effect seen in isotropic + noise simulations. This procedure can be used for a reliable determination of the signal in future surveys. However due to large systematics in current data the results obtained may not be reliable. We use the $R$ estimator results for this purpose. We define a quantity $d$ as
\begin{equation} 
    d = \frac{[\langle |P(\boldsymbol{\hat n})|^2\rangle_U - \langle |P(\boldsymbol{\hat n})|^2\rangle_D]_\text{max}}{\bar K},
\end{equation} 
where $\langle |P(\boldsymbol{\hat n})|^2\rangle_U$ and $\langle |P(\boldsymbol{\hat n})|^2\rangle_D$ are the upper and lower hemisphere averages as defined in Eq. \eqref{Eq:weightedP2} with the weights $w_{i,j}$ set to 1 and we define $\bar K$ as:
\begin{align}
    \bar K = &\left\{\int_U W^2(\boldsymbol{\hat n}) d\Omega \right\}^{-1} \int_U W^2(\boldsymbol{\hat n}) \cos \theta d\Omega \nonumber \\ &- \left\{\int_D W^2(\boldsymbol{\hat n})  d\Omega \right\}^{-1} \int_D W^2(\boldsymbol{\hat n}) \cos \theta d\Omega.
    \label{Eq:barKrel}
\end{align}
We estimate the quantity $d$ from our isotropic+noise FFP10 simulations and call this $d_N$. The quantity $\mathbf{d}_N$ represent the effective dipole arising due to random isotropic CMB and the noise and systematic residuals. The $d$ estimated from data, $\mathbf{d}_T$, is the dipole observed in the actual data. We subtract $\mathbf{d}_N$ from $\mathbf{d}_T$ vectorially to obtain the dipole modulation signal. We will only discuss the results for $\ell_\text{min}=10$. For the Commander map with \texttt{NSIDE}=64, we obtain $d_N=(1.65 \pm 0.83)\times 10^{-14}$ K$^2$. The corresponding mean direction is found to be $(21^\circ,32^\circ)$ in galactic coordinates. The resulting bias corrected values of the parameters are found to be $A = 0.52$ with preferred direction $(341^\circ,-28^\circ)$. 
For SMICA, we get $d_N=(1.46 \pm 0.74)\times 10^{-14}$ K$^2$ along $(40^\circ,37^\circ)$ and gives 
$A = 0.35$ with preferred direction $(224^\circ,-18^\circ)$. We get  $d_N=(1.57 \pm 0.73)\times 10^{-14}$ K$^2$ along $(38^\circ,44^\circ)$ for SEVEM, which gives us $A = 0.58$ with preferred direction $(340^\circ,-35^\circ)$. Finally, for NILC, we have $d_N=(4.19 \pm 1.41)\times 10^{-14}$ K$^2$ along $(1^\circ,3^\circ)$, with $A = 0.28$ with preferred direction $(224^\circ,-27^\circ)$. We note that the amplitude extracted after removal of an effective noise dipole from the FFP10 simulations, the amplitude is still large. One can also see that SMICA and NILC maps give comparable results, while the Commander and SEVEM are close to each other. Overall we find that a simple method of removing a noise dipole, estimated from the FFP10 simulations is still not sufficient to account for the noise-and-systematics bias present in the data. The systematics dipole removal discussed here is only an initial exploratory study and it needs further improvement which we will postpone to a future work. A more complete model for the noise correlations would probably be needed to effectively deal with the bias in the data.


\section{Conclusions}
We have developed a pixel based method to test for dipole modulation effect
in CMB polarization. The method is based on directly 
using the observed polarized flux $|P|^2 = Q^2+U^2$ and can be implemented
easily on masked sky. We have proposed a simple statistics in order to 
characterize the effect. The statistic quantifies the difference in the mean
value of $|P|^2$ in two hemispheres along some chosen direction. Alternatively
one can directly extract the dipole harmonic coefficients from data. We
have determined the performance of our statistics for a future CORE-like
mission. We find that if the effect is present in data at the level 
expected from the dipole modulation seen in CMB temperature, it may be
detectable at 2.7$\sigma$ level or better. The level of detection varies between 2.7$\sigma$ to 5.8$\sigma$ for $\pm1\sigma$ variation about the mean modulation amplitude. We also apply our method to
Planck 2018 CMB polarization data. It is well known that this data contains residual systematics 
bias and is not reliable for study of large scale isotropy. We find
that the dipole modulation is present in data in a direction which for many of the cases is very 
close to the galactic plane. The amplitude is found to be
relatively large. The results are consistent with the amplitude obtained from the FFP10 simulation for SMICA and NILC methods. But the results are larger than FFP10 simulation results for Commander and SEVEM, as indicated by the very small p-values. 
We are aware of a bias from the systematic residuals that is present in data. We subtract this bias vectorially from the dipole modulation seen in real data. The resulting dipole modulation parameters provide our best estimate for the effect. However the Planck 2018 data is not reliable for statistical isotropy testing and our estimator is sensitive to the bias arising from systematics. Despite the apparent significance of the Commander and SEVEM results, the disagreement among the different maps, the direction of modulation being close to the galactic plane and the limitations of our estimator in dealing with correlated noise, we are unable to claim a positive detection of dipole modulation signal in CMB polarization.

\section*{Acknowledgements}

Based on observations obtained with Planck (http://www.esa.int/Planck), an ESA science mission with instruments and contributions directly funded by ESA Member States, NASA, and Canada. The authors acknowledge funding from the Science and Engineering Research Board (SERB), Government of India, for  this research project. Some of the results in this paper have been derived using the HEALPix \citep{Gorski:2005} package. 




\bibliographystyle{mnras}
\bibliography{references} 


\bsp	
\label{lastpage}
\end{document}